\pdfoutput=1
\documentclass[11pt,twoside,a4paper,cmspaper,final,collab]{cms-tdr}
\def\svnVersion{32339:32423}\def\cmsCernNo{2010-088}\def\cmsCernDate{2011/01/17}\def\cmsMessage{Submitted to the Journal of High Energy Physics}

\begin{document}\cmsNoteHeader{BPH-10-007}
\hyphenation{env-iron-men-tal}
\hyphenation{had-ron-i-za-tion}
\hyphenation{cal-or-i-me-ter}
\hyphenation{de-vices}
\RCS$Revision: 32423 $
\RCS$HeadURL: svn+ssh://alverson@svn.cern.ch/reps/tdr2/papers/BPH-10-007/trunk/BPH-10-007.tex $
\RCS$Id: BPH-10-007.tex 32423 2011-01-18 17:26:58Z alverson $
%
%
%

\providecommand {\etal}{\mbox{et al.}\xspace} 
\providecommand {\ie}{\mbox{i.e.}\xspace}     
\providecommand {\eg}{\mbox{e.g.}\xspace}     
\providecommand {\etc}{\mbox{etc.}\xspace}     
\providecommand {\vs}{\mbox{\sl vs.}\xspace}      
\providecommand {\mdash}{\ensuremath{\mathrm{-}}} 

\providecommand {\Lone}{Level-1\xspace} 
\providecommand {\Ltwo}{Level-2\xspace}
\providecommand {\Lthree}{Level-3\xspace}

\providecommand{\ACERMC} {\textsc{AcerMC}\xspace}
\providecommand{\ALPGEN} {{\textsc{alpgen}}\xspace}
\providecommand{\CHARYBDIS} {{\textsc{charybdis}}\xspace}
\providecommand{\CMKIN} {\textsc{cmkin}\xspace}
\providecommand{\CMSIM} {{\textsc{cmsim}}\xspace}
\providecommand{\CMSSW} {{\textsc{cmssw}}\xspace}
\providecommand{\COBRA} {{\textsc{cobra}}\xspace}
\providecommand{\COCOA} {{\textsc{cocoa}}\xspace}
\providecommand{\COMPHEP} {\textsc{CompHEP}\xspace}
\providecommand{\EVTGEN} {{\textsc{evtgen}}\xspace}
\providecommand{\FAMOS} {{\textsc{famos}}\xspace}
\providecommand{\GARCON} {\textsc{garcon}\xspace}
\providecommand{\GARFIELD} {{\textsc{garfield}}\xspace}
\providecommand{\GEANE} {{\textsc{geane}}\xspace}
\providecommand{\GEANTfour} {{\textsc{geant4}}\xspace}
\providecommand{\GEANTthree} {{\textsc{geant3}}\xspace}
\providecommand{\GEANT} {{\textsc{geant}}\xspace}
\providecommand{\HDECAY} {\textsc{hdecay}\xspace}
\providecommand{\HERWIG} {{\textsc{herwig}}\xspace}
\providecommand{\HIGLU} {{\textsc{higlu}}\xspace}
\providecommand{\HIJING} {{\textsc{hijing}}\xspace}
\providecommand{\IGUANA} {\textsc{iguana}\xspace}
\providecommand{\ISAJET} {{\textsc{isajet}}\xspace}
\providecommand{\ISAPYTHIA} {{\textsc{isapythia}}\xspace}
\providecommand{\ISASUGRA} {{\textsc{isasugra}}\xspace}
\providecommand{\ISASUSY} {{\textsc{isasusy}}\xspace}
\providecommand{\ISAWIG} {{\textsc{isawig}}\xspace}
\providecommand{\MADGRAPH} {\textsc{MadGraph}\xspace}
\providecommand{\MCATNLO} {\textsc{mc@nlo}\xspace}
\providecommand{\MCFM} {\textsc{mcfm}\xspace}
\providecommand{\MILLEPEDE} {{\textsc{millepede}}\xspace}
\providecommand{\ORCA} {{\textsc{orca}}\xspace}
\providecommand{\OSCAR} {{\textsc{oscar}}\xspace}
\providecommand{\PHOTOS} {\textsc{photos}\xspace}
\providecommand{\PROSPINO} {\textsc{prospino}\xspace}
\providecommand{\PYTHIA} {{\textsc{pythia}}\xspace}
\providecommand{\SHERPA} {{\textsc{sherpa}}\xspace}
\providecommand{\TAUOLA} {\textsc{tauola}\xspace}
\providecommand{\TOPREX} {\textsc{TopReX}\xspace}
\providecommand{\XDAQ} {{\textsc{xdaq}}\xspace}

\providecommand {\DZERO}{D\O\xspace}     


\providecommand{\de}{\ensuremath{^\circ}}
\providecommand{\ten}[1]{\ensuremath{\times \text{10}^\text{#1}}}
\providecommand{\unit}[1]{\ensuremath{\text{\,#1}}\xspace}
\providecommand{\mum}{\ensuremath{\,\mu\text{m}}\xspace}
\providecommand{\micron}{\ensuremath{\,\mu\text{m}}\xspace}
\providecommand{\cm}{\ensuremath{\,\text{cm}}\xspace}
\providecommand{\mm}{\ensuremath{\,\text{mm}}\xspace}
\providecommand{\mus}{\ensuremath{\,\mu\text{s}}\xspace}
\providecommand{\keV}{\ensuremath{\,\text{ke\hspace{-.08em}V}}\xspace}
\providecommand{\MeV}{\ensuremath{\,\text{Me\hspace{-.08em}V}}\xspace}
\providecommand{\GeV}{\ensuremath{\,\text{Ge\hspace{-.08em}V}}\xspace}
\providecommand{\gev}{\GeV}
\providecommand{\TeV}{\ensuremath{\,\text{Te\hspace{-.08em}V}}\xspace}
\providecommand{\PeV}{\ensuremath{\,\text{Pe\hspace{-.08em}V}}\xspace}
\providecommand{\keVc}{\ensuremath{{\,\text{ke\hspace{-.08em}V\hspace{-0.16em}/\hspace{-0.08em}}c}}\xspace}
\providecommand{\MeVc}{\ensuremath{{\,\text{Me\hspace{-.08em}V\hspace{-0.16em}/\hspace{-0.08em}}c}}\xspace}
\providecommand{\GeVc}{\ensuremath{{\,\text{Ge\hspace{-.08em}V\hspace{-0.16em}/\hspace{-0.08em}}c}}\xspace}
\providecommand{\TeVc}{\ensuremath{{\,\text{Te\hspace{-.08em}V\hspace{-0.16em}/\hspace{-0.08em}}c}}\xspace}
\providecommand{\keVcc}{\ensuremath{{\,\text{ke\hspace{-.08em}V\hspace{-0.16em}/\hspace{-0.08em}}c^\text{2}}}\xspace}
\providecommand{\MeVcc}{\ensuremath{{\,\text{Me\hspace{-.08em}V\hspace{-0.16em}/\hspace{-0.08em}}c^\text{2}}}\xspace}
\providecommand{\GeVcc}{\ensuremath{{\,\text{Ge\hspace{-.08em}V\hspace{-0.16em}/\hspace{-0.08em}}c^\text{2}}}\xspace}
\providecommand{\TeVcc}{\ensuremath{{\,\text{Te\hspace{-.08em}V\hspace{-0.16em}/\hspace{-0.08em}}c^\text{2}}}\xspace}

\providecommand{\pbinv} {\mbox{\ensuremath{\,\text{pb}^\text{$-$1}}}\xspace}
\providecommand{\fbinv} {\mbox{\ensuremath{\,\text{fb}^\text{$-$1}}}\xspace}
\providecommand{\nbinv} {\mbox{\ensuremath{\,\text{nb}^\text{$-$1}}}\xspace}
\providecommand{\percms}{\ensuremath{\,\text{cm}^\text{$-$2}\,\text{s}^\text{$-$1}}\xspace}
\providecommand{\lumi}{\ensuremath{\mathcal{L}}\xspace}
\providecommand{\Lumi}{\ensuremath{\mathcal{L}}\xspace}
%
\providecommand{\LvLow}  {\ensuremath{\mathcal{L}=\text{10}^\text{32}\,\text{cm}^\text{$-$2}\,\text{s}^\text{$-$1}}\xspace}
\providecommand{\LLow}   {\ensuremath{\mathcal{L}=\text{10}^\text{33}\,\text{cm}^\text{$-$2}\,\text{s}^\text{$-$1}}\xspace}
\providecommand{\lowlumi}{\ensuremath{\mathcal{L}=\text{2}\times \text{10}^\text{33}\,\text{cm}^\text{$-$2}\,\text{s}^\text{$-$1}}\xspace}
\providecommand{\LMed}   {\ensuremath{\mathcal{L}=\text{2}\times \text{10}^\text{33}\,\text{cm}^\text{$-$2}\,\text{s}^\text{$-$1}}\xspace}
\providecommand{\LHigh}  {\ensuremath{\mathcal{L}=\text{10}^\text{34}\,\text{cm}^\text{$-$2}\,\text{s}^\text{$-$1}}\xspace}
\providecommand{\hilumi} {\ensuremath{\mathcal{L}=\text{10}^\text{34}\,\text{cm}^\text{$-$2}\,\text{s}^\text{$-$1}}\xspace}


\providecommand{\PT}{\ensuremath{p_{\mathrm{T}}}\xspace}
\providecommand{\pt}{\ensuremath{p_{\mathrm{T}}}\xspace}
\providecommand{\ET}{\ensuremath{E_{\mathrm{T}}}\xspace}
\providecommand{\HT}{\ensuremath{H_{\mathrm{T}}}\xspace}
\providecommand{\et}{\ensuremath{E_{\mathrm{T}}}\xspace}
\providecommand{\Em}{\ensuremath{E\hspace{-0.6em}/}\xspace}
\providecommand{\Pm}{\ensuremath{p\hspace{-0.5em}/}\xspace}
\providecommand{\PTm}{\ensuremath{{p}_\mathrm{T}\hspace{-1.02em}/}\xspace}
\providecommand{\PTslash}{\ensuremath{{p}_\mathrm{T}\hspace{-1.02em}/}\xspace}
\providecommand{\ETm}{\ensuremath{E_{\mathrm{T}}^{\text{miss}}}\xspace}
\providecommand{\MET}{\ETm}
\providecommand{\ETmiss}{\ETm}
\providecommand{\ETslash}{\ensuremath{E_{\mathrm{T}}\hspace{-1.1em}/}\xspace}
\providecommand{\VEtmiss}{\ensuremath{{\vec E}_{\mathrm{T}}^{\text{miss}}}\xspace}

\providecommand{\dd}[2]{\ensuremath{\frac{\mathrm{d} #1}{\mathrm{d} #2}}}
\providecommand{\ddinline}[2]{\ensuremath{\mathrm{d} #1/\mathrm{d} #2}}

\ifthenelse{\boolean{cms@italic}}{\newcommand{\cmsSymbolFace}{\relax}}{\newcommand{\cmsSymbolFace}{\mathrm}}

\providecommand{\zp}{\ensuremath{\cmsSymbolFace{Z}^\prime}\xspace}
\providecommand{\JPsi}{\ensuremath{\cmsSymbolFace{J}\hspace{-.08em}/\hspace{-.14em}\psi}\xspace}
\providecommand{\Z}{\ensuremath{\cmsSymbolFace{Z}}\xspace}
\providecommand{\ttbar}{\ensuremath{\cmsSymbolFace{t}\overline{\cmsSymbolFace{t}}}\xspace}

\newcommand{\cPgn}{\ensuremath{\nu}}
\newcommand{\cPJgy}{\JPsi}
\newcommand{\cPZ}{\Z}
\newcommand{\cPZpr}{\zp}


\providecommand{\AFB}{\ensuremath{A_\text{FB}}\xspace}
\providecommand{\wangle}{\ensuremath{\sin^{2}\theta_{\text{eff}}^\text{lept}(M^2_\Z)}\xspace}
\providecommand{\stat}{\ensuremath{\,\text{(stat.)}}\xspace}
\providecommand{\syst}{\ensuremath{\,\text{(syst.)}}\xspace}
\providecommand{\kt}{\ensuremath{k_{\mathrm{T}}}\xspace}

\providecommand{\BC}{\ensuremath{\mathrm{B_{c}}}\xspace}
\providecommand{\bbarc}{\ensuremath{\mathrm{\overline{b}c}}\xspace}
\providecommand{\bbbar}{\ensuremath{\mathrm{b\overline{b}}}\xspace}
\providecommand{\ccbar}{\ensuremath{\mathrm{c\overline{c}}}\xspace}
\providecommand{\bspsiphi}{\ensuremath{\mathrm{B_s} \to \JPsi\, \phi}\xspace}
\providecommand{\EE}{\ensuremath{\mathrm{e^+e^-}}\xspace}
\providecommand{\MM}{\ensuremath{\mu^+\mu^-}\xspace}
\providecommand{\TT}{\ensuremath{\tau^+\tau^-}\xspace}

\providecommand{\HGG}{\ensuremath{\mathrm{H}\to\gamma\gamma}}
\providecommand{\GAMJET}{\ensuremath{\gamma + \text{jet}}}
\providecommand{\PPTOJETS}{\ensuremath{\mathrm{pp}\to\text{jets}}}
\providecommand{\PPTOGG}{\ensuremath{\mathrm{pp}\to\gamma\gamma}}
\providecommand{\PPTOGAMJET}{\ensuremath{\mathrm{pp}\to\gamma + \mathrm{jet}}}
\providecommand{\MH}{\ensuremath{M_{\mathrm{H}}}}
\providecommand{\RNINE}{\ensuremath{R_\mathrm{9}}}
\providecommand{\DR}{\ensuremath{\Delta R}}

%

\providecommand{\ga}{\ensuremath{\gtrsim}}
\providecommand{\la}{\ensuremath{\lesssim}}
\providecommand{\swsq}{\ensuremath{\sin^2\theta_\cmsSymbolFace{W}}\xspace}
\providecommand{\cwsq}{\ensuremath{\cos^2\theta_\cmsSymbolFace{W}}\xspace}
\providecommand{\tanb}{\ensuremath{\tan\beta}\xspace}
\providecommand{\tanbsq}{\ensuremath{\tan^{2}\beta}\xspace}
\providecommand{\sidb}{\ensuremath{\sin 2\beta}\xspace}
\providecommand{\alpS}{\ensuremath{\alpha_S}\xspace}
\providecommand{\alpt}{\ensuremath{\tilde{\alpha}}\xspace}

\providecommand{\QL}{\ensuremath{\cmsSymbolFace{Q}_\cmsSymbolFace{L}}\xspace}
\providecommand{\sQ}{\ensuremath{\tilde{\cmsSymbolFace{Q}}}\xspace}
\providecommand{\sQL}{\ensuremath{\tilde{\cmsSymbolFace{Q}}_\cmsSymbolFace{L}}\xspace}
\providecommand{\ULC}{\ensuremath{\cmsSymbolFace{U}_\cmsSymbolFace{L}^\cmsSymbolFace{C}}\xspace}
\providecommand{\sUC}{\ensuremath{\tilde{\cmsSymbolFace{U}}^\cmsSymbolFace{C}}\xspace}
\providecommand{\sULC}{\ensuremath{\tilde{\cmsSymbolFace{U}}_\cmsSymbolFace{L}^\cmsSymbolFace{C}}\xspace}
\providecommand{\DLC}{\ensuremath{\cmsSymbolFace{D}_\cmsSymbolFace{L}^\cmsSymbolFace{C}}\xspace}
\providecommand{\sDC}{\ensuremath{\tilde{\cmsSymbolFace{D}}^\cmsSymbolFace{C}}\xspace}
\providecommand{\sDLC}{\ensuremath{\tilde{\cmsSymbolFace{D}}_\cmsSymbolFace{L}^\cmsSymbolFace{C}}\xspace}
\providecommand{\LL}{\ensuremath{\cmsSymbolFace{L}_\cmsSymbolFace{L}}\xspace}
\providecommand{\sL}{\ensuremath{\tilde{\cmsSymbolFace{L}}}\xspace}
\providecommand{\sLL}{\ensuremath{\tilde{\cmsSymbolFace{L}}_\cmsSymbolFace{L}}\xspace}
\providecommand{\ELC}{\ensuremath{\cmsSymbolFace{E}_\cmsSymbolFace{L}^\cmsSymbolFace{C}}\xspace}
\providecommand{\sEC}{\ensuremath{\tilde{\cmsSymbolFace{E}}^\cmsSymbolFace{C}}\xspace}
\providecommand{\sELC}{\ensuremath{\tilde{\cmsSymbolFace{E}}_\cmsSymbolFace{L}^\cmsSymbolFace{C}}\xspace}
\providecommand{\sEL}{\ensuremath{\tilde{\cmsSymbolFace{E}}_\cmsSymbolFace{L}}\xspace}
\providecommand{\sER}{\ensuremath{\tilde{\cmsSymbolFace{E}}_\cmsSymbolFace{R}}\xspace}
\providecommand{\sFer}{\ensuremath{\tilde{\cmsSymbolFace{f}}}\xspace}
\providecommand{\sQua}{\ensuremath{\tilde{\cmsSymbolFace{q}}}\xspace}
\providecommand{\sUp}{\ensuremath{\tilde{\cmsSymbolFace{u}}}\xspace}
\providecommand{\suL}{\ensuremath{\tilde{\cmsSymbolFace{u}}_\cmsSymbolFace{L}}\xspace}
\providecommand{\suR}{\ensuremath{\tilde{\cmsSymbolFace{u}}_\cmsSymbolFace{R}}\xspace}
\providecommand{\sDw}{\ensuremath{\tilde{\cmsSymbolFace{d}}}\xspace}
\providecommand{\sdL}{\ensuremath{\tilde{\cmsSymbolFace{d}}_\cmsSymbolFace{L}}\xspace}
\providecommand{\sdR}{\ensuremath{\tilde{\cmsSymbolFace{d}}_\cmsSymbolFace{R}}\xspace}
\providecommand{\sTop}{\ensuremath{\tilde{\cmsSymbolFace{t}}}\xspace}
\providecommand{\stL}{\ensuremath{\tilde{\cmsSymbolFace{t}}_\cmsSymbolFace{L}}\xspace}
\providecommand{\stR}{\ensuremath{\tilde{\cmsSymbolFace{t}}_\cmsSymbolFace{R}}\xspace}
\providecommand{\stone}{\ensuremath{\tilde{\cmsSymbolFace{t}}_1}\xspace}
\providecommand{\sttwo}{\ensuremath{\tilde{\cmsSymbolFace{t}}_2}\xspace}
\providecommand{\sBot}{\ensuremath{\tilde{\cmsSymbolFace{b}}}\xspace}
\providecommand{\sbL}{\ensuremath{\tilde{\cmsSymbolFace{b}}_\cmsSymbolFace{L}}\xspace}
\providecommand{\sbR}{\ensuremath{\tilde{\cmsSymbolFace{b}}_\cmsSymbolFace{R}}\xspace}
\providecommand{\sbone}{\ensuremath{\tilde{\cmsSymbolFace{b}}_1}\xspace}
\providecommand{\sbtwo}{\ensuremath{\tilde{\cmsSymbolFace{b}}_2}\xspace}
\providecommand{\sLep}{\ensuremath{\tilde{\cmsSymbolFace{l}}}\xspace}
\providecommand{\sLepC}{\ensuremath{\tilde{\cmsSymbolFace{l}}^\cmsSymbolFace{C}}\xspace}
\providecommand{\sEl}{\ensuremath{\tilde{\cmsSymbolFace{e}}}\xspace}
\providecommand{\sElC}{\ensuremath{\tilde{\cmsSymbolFace{e}}^\cmsSymbolFace{C}}\xspace}
\providecommand{\seL}{\ensuremath{\tilde{\cmsSymbolFace{e}}_\cmsSymbolFace{L}}\xspace}
\providecommand{\seR}{\ensuremath{\tilde{\cmsSymbolFace{e}}_\cmsSymbolFace{R}}\xspace}
\providecommand{\snL}{\ensuremath{\tilde{\nu}_L}\xspace}
\providecommand{\sMu}{\ensuremath{\tilde{\mu}}\xspace}
\providecommand{\sNu}{\ensuremath{\tilde{\nu}}\xspace}
\providecommand{\sTau}{\ensuremath{\tilde{\tau}}\xspace}
\providecommand{\Glu}{\ensuremath{\cmsSymbolFace{g}}\xspace}
\providecommand{\sGlu}{\ensuremath{\tilde{\cmsSymbolFace{g}}}\xspace}
\providecommand{\Wpm}{\ensuremath{\cmsSymbolFace{W}^{\pm}}\xspace}
\providecommand{\sWpm}{\ensuremath{\tilde{\cmsSymbolFace{W}}^{\pm}}\xspace}
\providecommand{\Wz}{\ensuremath{\cmsSymbolFace{W}^{0}}\xspace}
\providecommand{\sWz}{\ensuremath{\tilde{\cmsSymbolFace{W}}^{0}}\xspace}
\providecommand{\sWino}{\ensuremath{\tilde{\cmsSymbolFace{W}}}\xspace}
\providecommand{\Bz}{\ensuremath{\cmsSymbolFace{B}^{0}}\xspace}
\providecommand{\sBz}{\ensuremath{\tilde{\cmsSymbolFace{B}}^{0}}\xspace}
\providecommand{\sBino}{\ensuremath{\tilde{\cmsSymbolFace{B}}}\xspace}
\providecommand{\Zz}{\ensuremath{\cmsSymbolFace{Z}^{0}}\xspace}
\providecommand{\sZino}{\ensuremath{\tilde{\cmsSymbolFace{Z}}^{0}}\xspace}
\providecommand{\sGam}{\ensuremath{\tilde{\gamma}}\xspace}
\providecommand{\chiz}{\ensuremath{\tilde{\chi}^{0}}\xspace}
\providecommand{\chip}{\ensuremath{\tilde{\chi}^{+}}\xspace}
\providecommand{\chim}{\ensuremath{\tilde{\chi}^{-}}\xspace}
\providecommand{\chipm}{\ensuremath{\tilde{\chi}^{\pm}}\xspace}
\providecommand{\Hone}{\ensuremath{\cmsSymbolFace{H}_\cmsSymbolFace{d}}\xspace}
\providecommand{\sHone}{\ensuremath{\tilde{\cmsSymbolFace{H}}_\cmsSymbolFace{d}}\xspace}
\providecommand{\Htwo}{\ensuremath{\cmsSymbolFace{H}_\cmsSymbolFace{u}}\xspace}
\providecommand{\sHtwo}{\ensuremath{\tilde{\cmsSymbolFace{H}}_\cmsSymbolFace{u}}\xspace}
\providecommand{\sHig}{\ensuremath{\tilde{\cmsSymbolFace{H}}}\xspace}
\providecommand{\sHa}{\ensuremath{\tilde{\cmsSymbolFace{H}}_\cmsSymbolFace{a}}\xspace}
\providecommand{\sHb}{\ensuremath{\tilde{\cmsSymbolFace{H}}_\cmsSymbolFace{b}}\xspace}
\providecommand{\sHpm}{\ensuremath{\tilde{\cmsSymbolFace{H}}^{\pm}}\xspace}
\providecommand{\hz}{\ensuremath{\cmsSymbolFace{h}^{0}}\xspace}
\providecommand{\Hz}{\ensuremath{\cmsSymbolFace{H}^{0}}\xspace}
\providecommand{\Az}{\ensuremath{\cmsSymbolFace{A}^{0}}\xspace}
\providecommand{\Hpm}{\ensuremath{\cmsSymbolFace{H}^{\pm}}\xspace}
\providecommand{\sGra}{\ensuremath{\tilde{\cmsSymbolFace{G}}}\xspace}
\providecommand{\mtil}{\ensuremath{\tilde{m}}\xspace}
\providecommand{\rpv}{\ensuremath{\rlap{\kern.2em/}R}\xspace}
\providecommand{\LLE}{\ensuremath{LL\bar{E}}\xspace}
\providecommand{\LQD}{\ensuremath{LQ\bar{D}}\xspace}
\providecommand{\UDD}{\ensuremath{\overline{UDD}}\xspace}
\providecommand{\Lam}{\ensuremath{\lambda}\xspace}
\providecommand{\Lamp}{\ensuremath{\lambda'}\xspace}
\providecommand{\Lampp}{\ensuremath{\lambda''}\xspace}
\providecommand{\spinbd}[2]{\ensuremath{\bar{#1}_{\dot{#2}}}\xspace}

\providecommand{\MD}{\ensuremath{{M_\mathrm{D}}}\xspace}
\providecommand{\Mpl}{\ensuremath{{M_\mathrm{Pl}}}\xspace}
\providecommand{\Rinv} {\ensuremath{{R}^{-1}}\xspace} 
\cmsNoteHeader{BPH-10-007} 
\title{Inclusive b-hadron production cross section with muons in pp collisions at $\sqrt{s} = 7\,\mathrm{TeV}$}

\date{\today}

\def\rmm   {\ensuremath{\Delta R(\mu\mu)}}
\def\mll   {\ensuremath{m_{\mu\mu}}}
\def\dof   {\ensuremath\mathrm{dof}}
\def\ket#1        {\ensuremath|{#1}\rangle}
\def\bra#1        {\ensuremath\langle{#1}|}
\def\braket#1#2   {\ensuremath\langle{#1}|{#2}\rangle}
\def\tfi#1#2#3    {\ensuremath\langle{#1}|{#2}|{#3}\rangle}
\def\vtwo#1#2     {\ensuremath\left(\begin{array}{c}{#1}\\{#2}\end{array}\right)}
\def\vthree#1#2#3 {\ensuremath\left(\begin{array}{c}{#1}\\{#2}\\{#3}\end{array}\right)}
\def\me           {\ensuremath\mathcal{M}}
\def\ame          {\ensuremath|\mathcal{M}|^2}
\def\asme         {\ensuremath\overline{|\mathcal{M}|^2}}

\def\psib         {\ensuremath\overline{\psi}}

\newcommand{\mat}[2][cccccccccccccccccccccccccc]{\left(
   \begin{array}{#1}
    #2\\
   \end{array}
  \right)
}

\def\vvA        {\ensuremath\left(\begin{matrix} 0 \\ 0 \end{matrix}\right)}
\def\vvB        {\ensuremath\left(\begin{matrix} 1 \\ 0 \end{matrix}\right)}
\def\vvC        {\ensuremath\left(\begin{matrix} 0 \\ 1 \end{matrix}\right)}
\def\vvD        {\ensuremath\left(\begin{matrix} 1 \\ 1 \end{matrix}\right)}

\def\vvvA        {\ensuremath\left(\begin{matrix} 0 \\ 0 \\ 0\end{matrix}\right)}
\def\vvvB        {\ensuremath\left(\begin{matrix} 1 \\ 0 \\ 0\end{matrix}\right)}
\def\vvvC        {\ensuremath\left(\begin{matrix} 0 \\ 1 \\ 0\end{matrix}\right)}
\def\vvvD        {\ensuremath\left(\begin{matrix} 0 \\ 0 \\ 1\end{matrix}\right)}
\def\cpt   {\ensuremath{C\kern-0.2em P\kern-0.1em T}}
\def\cp    {\ensuremath{C\kern-0.2em P}}
\def\cpv   {\ensuremath{C\kern-0.2em P\kern-1.0em / }}
\def\CPV   {\cp-violation}
\def\CPTV  {\cpt-violation}
\def\bfsx  {$B$-physics}
\def\ETm   {\ensuremath{E_T\kern-1.2em/\kern0.6em}}
\def\ET    {\ensuremath{E_T}}
\def\kT    {\ensuremath{k_T}}
\def\ptm   {\ensuremath{p_\perp\kern-1.1em/\kern0.5em}}
\def\pvecm {\ensuremath{\vec{p} \kern-0.4em/\kern0.1em}}
\def\pvec  {\ensuremath{\vec{p}}}

\def\dsj   {\ensuremath{D_{sJ}}}
\def\vxb   {\ensuremath{|V_{xb}|}}
\def\vud   {\ensuremath{|V_{ud}|}}
\def\vus   {\ensuremath{|V_{us}|}}
\def\vub   {\ensuremath{|V_{ub}|}}
\def\vcd   {\ensuremath{|V_{cd}|}}
\def\vcs   {\ensuremath{|V_{cs}|}}
\def\vcb   {\ensuremath{|V_{cb}|}}
\def\vtd   {\ensuremath{|V_{td}|}}
\def\vts   {\ensuremath{|V_{ts}|}}
\def\vtb   {\ensuremath{|V_{tb}|}}

\def\deltam{\ensuremath{\delta m}}
\def\dm    {\ensuremath{\Delta m}}
\def\dt    {\ensuremath{\Delta t}}
\def\dg    {\ensuremath{\Delta \gamma}}
\def\dG    {\ensuremath{\Delta \Gamma}}
\def\dmt   {\ensuremath{\Delta mt}}
\def\dmdt  {\ensuremath{\Delta m \Delta t}}
\def\dms   {\ensuremath{\Delta m_s}}
\def\dmst {\ensuremath{\Delta m_s t}}
\def\dmm   {\ensuremath{\Delta m^2}}
\def\TBY   {\ensuremath{\theta_{\Bz, D^*\ell}}}

\def\de    {\ensuremath{\Delta E}}
\def\mes   {\ensuremath{m_{ES}}}

\def\msd{\ensuremath{\overline{m}_D^2}}
\def\lbar{\ensuremath{\overline{\Lambda}}}
\def\lone{\ensuremath{\lambda_1}}
\def\ltwo{\ensuremath{\lambda_2}}

\def\MUP   {\ensuremath{\mu_\pi^2}}
\def\MUG   {\ensuremath{\mu_G^2}}
\def\RHOD  {\ensuremath{\rho_D^3}}
\def\RHOLS {\ensuremath{\rho_{LS}^3}}

\def\cbf {\ensuremath{{\cal B}}}
\def\clu {\ensuremath{{\cal L}}}
\def\cor {\ensuremath{{\cal O}}}
\def\mmiss{\ensuremath{{m_{miss}^2}}}
\def\rusl{\ensuremath{{R_{u}}}}
\def\mh{\ensuremath{{m_{had}}}}
\def\mmxx {\ensuremath{\langle m_X^2 \rangle~}}
\def\mmx {\ensuremath{\langle m_X \rangle~}}
\def\mxqq{\ensuremath{(m_X, Q^2)}}
\def\mX{\ensuremath{{m_X}}}
\def\mx{\ensuremath{{m_X}}}
\def\mxcut{\ensuremath{{m_X^{cut}}}}
\def\pstar{\ensuremath{{p^*}}}
\def\qtot{\ensuremath{{Q_{tot}}}}
\def\mt{\ensuremath{{m_\perp}}}

\def\pslash{\ensuremath{{p\kern-0.45em /}}}
\def\pvecslash{\ensuremath{{\vec{p} \kern-0.45em /}}}
\def\meanmxx   {\ensuremath{\langle m_X^2 \rangle}}
\def\meanmx    {\ensuremath{\langle m_{X} \rangle}}
\def\mean#1    {\ensuremath{\langle #1 \rangle}}
\def\Bpilnu    {\ensuremath{\Bb\to \pi\ell\nub}}
\def\Bmpilnu   {\ensuremath{\Bm\to \pi^0\ell^-\nub}}
\def\Betalnu   {\ensuremath{\Bb\to \eta\ell\nub}}
\def\Brholnu   {\ensuremath{\Bb\to \rho\ell\nub}}
\def\Bmrholnu  {\ensuremath{\Bb\to \rho^0\ell^-\nub}}
\def\Bomegalnu {\ensuremath{\Bb\to \omega\ell\nub}}
\def\Brhoenu   {\ensuremath{\Bb\to \rho e\nub}}
\def\Bzrhoenu  {\ensuremath{\Bz\to \rho^- e^+\nub}}

\def\Bdlnu     {\ensuremath{\Bb\to D\ell\nub}}
\def\Bdstarlnu {\ensuremath{\Bb\to \Dstar \ell \nub}}
\def\Bzdstarlnu {\ensuremath{\Bzb\to \Dstarp \ell^- \nub}}
\def\Bzdstarenu {\ensuremath{\Bzb\to \Dstarp e^- \nub}}

\def\bll     {\ensuremath{\Bz\to \ell^+\ell^-}}
\def\bee     {\ensuremath{\Bz\to e^+e^-}}
\def\bmm     {\ensuremath{\Bz\to \mu^+\mu^-}}
\def\bem     {\ensuremath{\Bz\to e^\pm\mu^\mp}}
\def\btt     {\ensuremath{\Bs\to \tau^+\tau^-}}
\def\bmt     {\ensuremath{\Bs\to \mu^\pm\tau^\mp}}
\def\bmn     {\ensuremath{B^+\to \mu^+\nu_\mu}}
\def\btn     {\ensuremath{B^+\to \tau^+\nu_\tau}}
\def\bln     {\ensuremath{B^+\to \ell^+\nu_\ell}}
\def\bgen    {\ensuremath{B^-\to \gamma e\nub}}
\def\bgee    {\ensuremath{B^-\to \gamma e^+e^-}}
\def\bgg     {\ensuremath{B^-\to \gamma \gamma}}

\def\jpsitomu {\ensuremath{\jpsi\to \mu^+\mu^-}}
\def\bdmm     {\ensuremath{B^0\to  \mu^+\mu^-}}
\def\bdpipi   {\ensuremath{B^0\to  \pip\pim}}
\def\bdpik    {\ensuremath{B^0\to  \Kp\pim}}
\def\bdpimunu {\ensuremath{B^0\to  \pim\mup\nu}}
\def\bdmumupz {\ensuremath{B^0\to  \mup\mun\piz}}
\def\lbppi    {\ensuremath{\Lambda_b\to \pim p}}
\def\lbpk     {\ensuremath{\Lambda_b\to K^- p}}
\def\butrmunu {\ensuremath{B^+\to \mup\mun\mup \nu}}
\def\bctrmunu {\ensuremath{B_c\to \mup\mun\mup \nu}}
\def\bcpsimunu{\ensuremath{B_c\to \jpsi(\to\mup\mun)\mup \nu}}

\def\bdmm     {\ensuremath{B^0  \to \mu^+\mu^-}}
\def\bsmm     {\ensuremath{B^0_s\to \mu^+\mu^-}}
\def\bskk     {\ensuremath{B^0_s\to K^+K^-}}
\def\bspipi   {\ensuremath{B^0_s\to \pip\pim}}
\def\bspik    {\ensuremath{B^0_s\to K^-\pip}}
\def\bskmunu  {\ensuremath{B^0_s\to K^-\mup\nu}}
\def\bsmumug  {\ensuremath{B^0_s\to \mup\mun\gamma}}
\def\bsmmg    {\ensuremath{B^0_s\to \mu^+\mu^-\gamma}}
\def\bdll     {\ensuremath{B^0\to \ell^+\ell^-}}
\def\bsll     {\ensuremath{B^0_s\to \ell^+\ell^-}}
\def\bstt     {\ensuremath{B^0_s\to \tau^+\tau^-}}
\def\bdmt     {\ensuremath{B^0  \to \mu^\pm\tau^\mp}}
\def\bsmt     {\ensuremath{B^0_s\to \mu^\pm\tau^\mp}}
\def\bsdmm    {\ensuremath{B^0_{s (d)}\to \mu^+\mu^-}}
\def\bsdll    {\ensuremath{B^0_{s (d)}\to \ell^+\ell^-}}
\def\bsdtt    {\ensuremath{B^0_{s (d)}\to \tau^+\tau^-}}
\def\tmmm     {\ensuremath{\tau\to \mu\mu\mu}}

\def\hbb      {\ensuremath{H\to \bbbar}}
\def\htt      {\ensuremath{H\to \taup\taum}}
\def\ttH      {\ensuremath{\ttbar H}}

\def\meg      {\ensuremath{\mu\to e\gamma}}
\def\meee     {\ensuremath{\mu\to eee}}
\def\pgg      {\ensuremath{\piz\to \g\g}}

\def\bsg     {\ensuremath{b\to s\gamma}}
\def\bulnu   {\ensuremath{b\to u\ell\nub}}
\def\bclnu   {\ensuremath{b\to c\ell\nub}}
\def\bcenu   {\ensuremath{b\to c e\nub}}
\def\buenu   {\ensuremath{b\to u e\nub}}

\def\bxlnu   {\ensuremath{b\to X\ell^-\nub}}
\def\Bxenu   {\ensuremath{\Bb\to Xe^-\nu}}

\newcommand {\Bxlnu}{\ensuremath{\Bb \rightarrow X \ell \bar{\nu}}}
\newcommand {\Bxclnu}{\ensuremath{\Bb \rightarrow X_c \ell \bar{\nu}}}
\newcommand {\Bxulnu}{\ensuremath{\Bb \rightarrow X_u \ell \bar{\nu}}}
\def\Bpxenu {\ensuremath{\Bp\to Xe^+\nu}}
\def\Bzxenu {\ensuremath{\Bz\to Xe^+\nu}}

\def\Bulnu   {\ensuremath{\Bb\to X_{u}\ell\nub}}
\def\Buenu   {\ensuremath{\Bb\to X_{u} e\nub}}
\def\Bclnu   {\ensuremath{\Bb\to X_{c}\ell\nub}}
\def\Bxulnu  {\ensuremath{\Bb\to X_{u}\ell\nub}}
\def\Bxuenu  {\ensuremath{\Bb\to X_{u} e\nub}}
\def\Bxclnu  {\ensuremath{\Bb\to X_{c}\ell\nub}}

\def\bfactory  {{{\sl B}-factory}}
\def\bfactories{{{\sl B}-factories}}
\def\bFactory  {{{\sl B}-Factory}}
\def\breco     {\ensuremath{B_{reco}}}
\def\btag      {\ensuremath{B_{tag}}}
\def\bdecay    {{$B$-decay}}
\def\bDecay    {{$B$-Decay}}
\def\bdecays   {{$B$-decays}}
\def\bDecays   {{$B$-Decays}}
\def\bhadron   {{$b$-hadron}}
\def\bhadrons  {{$B$-hadrons}}
\def\bmeson    {{$B$-meson}}
\def\bmesons   {{$B$-mesons}}
\def\bquark    {{$b$-quark}}
\def\bquarks   {{$b$-quarks}}
\def\bphysics  {{$b$-physics}}
\def\Bphysics  {{$B$-physics}}
\def\ie   {{\it i.e.}}
\def\cf   {{\it cf.}}
\def\eg   {{\it e.g.}}
\def\etal {{\it et~al.}}
\def\etc  {{\it etc.}}
\def\rtr    {{$\red\triangleright\black$}}
\def\barrow {{$\blue\to\black$}}
\def\bpoint {{$\blue\bullet\black$}}
\def\npoint {{\phantom{$ \blue\bullet\black$}}}

\newcommand\bfac   {$B$-Factories}

\newcommand\bu   {\ensuremath{b\to u}}
\newcommand\bc   {\ensuremath{b\to c}}

\newcommand\islbcd {inclusive semileptonic $B\to c\ell\nu$}
\newcommand\islbud {inclusive semileptonic $B\to u\ell\nu$}
\def\tg     {\ensuremath {\theta^{*}_T}}
\def\ctg     {\ensuremath {\cos{\tg}}}
\def\cth     {\ensuremath {\cos{\theta_{H}}}}
\def\cthe    {\ensuremath {\cos{\theta_{H\,\eta'}}}}
\def\cthr    {\ensuremath {\cos{\theta_{H\,\rho}}}}
\def\ctb     {\ensuremath {\cos{\theta^{*}_{B}}}}
\def\ebeam     {\ensuremath {E^{*}_{b}}}
\def\egcms     {\ensuremath {E^{*}_{\gamma}}}
\def\mkpi      {\ensuremath {M_{\Kp \pim}}}
\let\emi\en
\def\electron   {\ensuremath{e}\xspace}
\def\en         {\ensuremath{e^-}\xspace}   
\def\ep         {\ensuremath{e^+}\xspace}
\def\epm        {\ensuremath{e^\pm}\xspace}
\def\epem       {\ensuremath{e^+e^-}\xspace}
\def\ee         {\ensuremath{e^-e^-}\xspace}

\def\mmu        {\ensuremath{\mu}\xspace}
\def\mup        {\ensuremath{\mu^+}\xspace}
\def\mun        {\ensuremath{\mu^-}\xspace} 
\def\mumu       {\ensuremath{\mu^+\mu^-}\xspace}
\def\mtau       {\ensuremath{\tau}\xspace}

\def\taup       {\ensuremath{\tau^+}\xspace}
\def\taum       {\ensuremath{\tau^-}\xspace}
\def\tautau     {\ensuremath{\tau^+\tau^-}\xspace}

\def\ellm       {\ensuremath{\ell^-}\xspace}
\def\ellp       {\ensuremath{\ell^+}\xspace}
\def\ellell     {\ensuremath{\ell^+ \ell^-}\xspace}

\def\ellb        {\ensuremath{\bar{\ell}}\xspace}
\def\nub        {\ensuremath{\bar{\nu}}\xspace}
\def\nunub      {\ensuremath{\nu{\bar{\nu}}}\xspace}
\def\nunub      {\ensuremath{\nu{\bar{\nu}}}\xspace}
\def\nue        {\ensuremath{\nu_e}\xspace}
\def\nueb       {\ensuremath{\nub_e}\xspace}
\def\nuenueb    {\ensuremath{\nue\nueb}\xspace}
\def\num        {\ensuremath{\nu_\mu}\xspace}
\def\numb       {\ensuremath{\nub_\mu}\xspace}
\def\numnumb    {\ensuremath{\num\numb}\xspace}
\def\nut        {\ensuremath{\nu_\tau}\xspace}
\def\nutb       {\ensuremath{\nub_\tau}\xspace}
\def\nutnutb    {\ensuremath{\nut\nutb}\xspace}
\def\nul        {\ensuremath{\nu_\ell}\xspace}
\def\nulb       {\ensuremath{\nub_\ell}\xspace}
\def\nulnulb    {\ensuremath{\nul\nulb}\xspace}

\def\g     {\ensuremath{\gamma}\xspace}
\def\gaga  {\ensuremath{\gamma\gamma}\xspace}  
\def\ggstar{\ensuremath{\gamma\gamma^*}\xspace}

\def\ega    {\ensuremath{e\gamma}\xspace}
\def\game   {\ensuremath{\gamma e^-}\xspace}
\def\epemg  {\ensuremath{e^+e^-\gamma}\xspace}

\def\H      {\ensuremath{H^0}\xspace}
\def\Hp     {\ensuremath{H^+}\xspace}
\def\Hm     {\ensuremath{H^-}\xspace}
\def\Hpm    {\ensuremath{H^\pm}\xspace}
\def\W      {\ensuremath{W}\xspace}
\def\Wp     {\ensuremath{W^+}\xspace}
\def\Wm     {\ensuremath{W^-}\xspace}
\def\Wpm    {\ensuremath{W^\pm}\xspace}
\def\Z      {\ensuremath{Z^0}\xspace}

\def\q     {\ensuremath{q}\xspace}
\def\qbar  {\ensuremath{\overline q}\xspace}
\def\Qbar  {\ensuremath{\overline Q}\xspace}
\def\ffbar {\ensuremath{f\overline f}\xspace}
\def\qqbar {\ensuremath{q\overline q}\xspace}
\def\QQbar {\ensuremath{Q\overline Q}\xspace}
\def\u     {\ensuremath{u}\xspace}
\def\ubar  {\ensuremath{\overline u}\xspace}
\def\uubar {\ensuremath{u\overline u}\xspace}
\def\d     {\ensuremath{d}\xspace}
\def\dbar  {\ensuremath{\overline d}\xspace}
\def\ddbar {\ensuremath{d\overline d}\xspace}
\def\s     {\ensuremath{s}\xspace}
\def\sbar  {\ensuremath{\overline s}\xspace}
\def\ssbar {\ensuremath{s\overline s}\xspace}
\def\c     {\ensuremath{c}\xspace}
\def\cbar  {\ensuremath{\overline c}\xspace}
\def\ccbar {\ensuremath{c\overline c}\xspace}
\def\b     {\ensuremath{b}\xspace}
\def\bbar  {\ensuremath{\overline b}\xspace}
\def\bbbar {\ensuremath{b\overline b}\xspace}
\def\t     {\ensuremath{t}\xspace}
\def\tbar  {\ensuremath{\overline t}\xspace}
\def\tbar  {\ensuremath{\overline t}\xspace}
\def\ttbar {\ensuremath{t\overline t}\xspace}
\def\pbar  {\ensuremath{\overline p}\xspace}
\def\ppbar {\ensuremath{p\overline p}\xspace}

\def\piz   {\ensuremath{\pi^0}\xspace}
\def\pizs  {\ensuremath{\pi^0\mbox\,\rm{s}}\xspace}
\def\ppz   {\ensuremath{\pi^0\pi^0}\xspace}
\def\pip   {\ensuremath{\pi^+}\xspace}
\def\pim   {\ensuremath{\pi^-}\xspace}
\def\pipi  {\ensuremath{\pi^+\pi^-}\xspace}
\def\pipm  {\ensuremath{\pi^\pm}\xspace}
\def\pimp  {\ensuremath{\pi^\mp}\xspace}

\def\kaon  {\ensuremath{K}\xspace}
\def\Kbar  {\kern 0.2em\bar{\kern -0.2em K}{}\xspace}
\def\Kb    {\ensuremath{\Kbar}\xspace}
\def\Kz    {\ensuremath{K^0}\xspace}
\def\Kzb   {\ensuremath{\Kbar^0}\xspace}
\def\KzKzb {\ensuremath{\Kz \kern -0.16em \Kzb}\xspace}
\def\Kp    {\ensuremath{K^+}\xspace}
\def\Km    {\ensuremath{K^-}\xspace}
\def\Kpm   {\ensuremath{K^\pm}\xspace}
\def\Kmp   {\ensuremath{K^\mp}\xspace}
\def\KpKm  {\ensuremath{\Kp \kern -0.16em \Km}\xspace}
\def\KS    {\ensuremath{K^0_{\scriptscriptstyle S}}\xspace}
\def\KL    {\ensuremath{K^0_{\scriptscriptstyle L}}\xspace}
\def\Kstarz  {\ensuremath{K^{*0}}\xspace}
\def\Kstarzb {\ensuremath{\Kbar^{*0}}\xspace}
\def\Kstar   {\ensuremath{K^*}\xspace}
\def\Kstarb  {\ensuremath{\Kbar^*}\xspace}
\def\Kstarp  {\ensuremath{K^{*+}}\xspace}
\def\Kstarm  {\ensuremath{K^{*-}}\xspace}
\def\Kstarpm {\ensuremath{K^{*\pm}}\xspace}
\def\Kstarmp {\ensuremath{K^{*\mp}}\xspace}

\newcommand{\etapr}{\ensuremath{\eta^{\prime}}\xspace}
\def\Dbar    {\kern 0.2em\bar{\kern -0.2em D}{}\xspace}
\def\Db      {\ensuremath{\Dbar}\xspace}
\def\Dz      {\ensuremath{D^0}\xspace}
\def\Dzb     {\ensuremath{\Dbar^0}\xspace}
\def\DzDzb   {\ensuremath{\Dz {\kern -0.16em \Dzb}}\xspace}
\def\Dp      {\ensuremath{D^+}\xspace}
\def\Dm      {\ensuremath{D^-}\xspace}
\def\Dpm     {\ensuremath{D^\pm}\xspace}
\def\Dmp     {\ensuremath{D^\mp}\xspace}
\def\DpDm    {\ensuremath{\Dp {\kern -0.16em \Dm}}\xspace}
\def\Dstar   {\ensuremath{D^*}\xspace}
\def\Dstarb  {\ensuremath{\Dbar^*}\xspace}
\def\Dstarz  {\ensuremath{D^{*0}}\xspace}
\def\Dstarzb {\ensuremath{\Dbar^{*0}}\xspace}
\def\Dstarp  {\ensuremath{D^{*+}}\xspace}
\def\Dstarm  {\ensuremath{D^{*-}}\xspace}
\def\Dstarpm {\ensuremath{D^{*\pm}}\xspace}
\def\Dstarmp {\ensuremath{D^{*\mp}}\xspace}
\def\Ds      {\ensuremath{D^+_s}\xspace}
\def\Dsb     {\ensuremath{\Dbar^+_s}\xspace}
\def\Dss     {\ensuremath{D^{*+}_s}\xspace}
\newcommand{\dstr}{\ensuremath{\Dstar}\xspace}
\newcommand{\dstrstr}{\ensuremath{D^{**}}\xspace}
\newcommand{\dsp}{\ensuremath{\Dstarp}\xspace}
\newcommand{\dsm}{\ensuremath{\Dstarm}\xspace}
\newcommand{\dsz}{\ensuremath{\Dstarz}\xspace}

\def\B       {\ensuremath{B}\xspace}
\def\Bbar    {\kern 0.18em\bar{\kern -0.18em B}{}\xspace}
\def\Bb      {\ensuremath{\Bbar}\xspace}
\def\BB      {\ensuremath{B\Bbar}\xspace}
\def\Bz      {\ensuremath{B^0}\xspace}
\def\Bzb     {\ensuremath{\Bbar^0}\xspace}
\def\BzBzb   {\ensuremath{\Bz {\kern -0.16em \Bzb}}\xspace}
\def\BsBsb   {\ensuremath{\Bs {\kern -0.16em \Bsb}}\xspace}
\def\Bu      {\ensuremath{B^+}\xspace}
\def\Bub     {\ensuremath{B^-}\xspace}
\def\Bp      {\ensuremath{\Bu}\xspace}
\def\Bm      {\ensuremath{\Bub}\xspace}
\def\Bpm     {\ensuremath{B^\pm}\xspace}
\def\Bmp     {\ensuremath{B^\mp}\xspace}
\def\BpBm    {\ensuremath{\Bu {\kern -0.16em \Bub}}\xspace}
\def\Bd      {\ensuremath{B^0_d}\xspace}
\def\Bs      {\ensuremath{B^0_s}\xspace}
\def\Bc      {\ensuremath{B^+_c}\xspace}
\def\Bsb     {\ensuremath{\Bzb_s}\xspace}
\def\Nz      {\ensuremath{M^0}\xspace}
\def\Nbar    {\kern 0.18em\bar{\kern -0.18em M}{}}
\def\Nzb     {\ensuremath{\Nbar^0}}
\def\NzNzb   {\ensuremath{\Nz {\kern -0.16em \Nzb}}}
\def\Nh      {\ensuremath{M_H}\xspace}
\def\Nl      {\ensuremath{M_L}\xspace}
\def\Nphys   {\ensuremath{\Nz_{phys}(t)}\xspace}
\def\Nbphys  {\ensuremath{\Nzb_{phys}(t)}\xspace}
\def\gh      {\ensuremath{\gamma_H}\xspace}
\def\gl      {\ensuremath{\gamma_L}\xspace}

\def\Bzd     {\ensuremath{B_d^0}\xspace}
\def\Bzs     {\ensuremath{B_s^0}\xspace}
\def\Bsd     {\ensuremath{B_{s(d)}}\xspace}

\def\jpsi     {\ensuremath{{J\mskip -3mu/\mskip -2mu\psi\mskip 2mu}}\xspace}
\def\psitwos  {\ensuremath{\psi{(2S)}}\xspace}
\def\psiprpr  {\ensuremath{\psi(3770)}\xspace}
\def\etac     {\ensuremath{\eta_c}\xspace}
\def\chiczero {\ensuremath{\chi_{c0}}\xspace}
\def\chicone  {\ensuremath{\chi_{c1}}\xspace}
\def\chictwo  {\ensuremath{\chi_{c2}}\xspace}
\mathchardef\Upsilon="7107
\def\Y#1S{\ensuremath{\Upsilon{(#1S)}}\xspace}
\def\OneS  {\Y1S}
\def\TwoS  {\Y2S}
\def\ThreeS{\Y3S}
\def\FourS {\Y4S}
\def\FiveS {\Y5S}
\def\chic#1{\ensuremath{\chi_{c#1}}\xspace} 

\def\proton      {\ensuremath{p}\xspace}
\def\antiproton  {\ensuremath{\overline p}\xspace}
\def\neutron     {\ensuremath{n}\xspace}
\def\antineutron {\ensuremath{\overline n}\xspace}

\mathchardef\Deltares="7101
\mathchardef\Xi="7104
\mathchardef\Lambda="7103
\mathchardef\Sigma="7106
\mathchardef\Omega="710A
\def\Deltabar{\kern 0.25em\overline{\kern -0.25em \Deltares}{}\xspace}
\def\Lbar{\kern 0.2em\overline{\kern -0.2em\Lambda\kern 0.05em}\kern-0.05em{}\xspace}
\def\Sigbar{\kern 0.2em\overline{\kern -0.2em \Sigma}{}\xspace}
\def\Xibar{\kern 0.2em\overline{\kern -0.2em \Xi}{}\xspace}
\def\Obar{\kern 0.2em\overline{\kern -0.2em \Omega}{}\xspace}
\def\Xb{\kern 0.2em\overline{\kern -0.2em X}{}\xspace}

\def\X {\ensuremath{X}\xspace}

\def\BR         {{\ensuremath{\cal B}\xspace}}
\def\BRtauptoe  {\ensuremath{\BR(\taup \to \ep)}\xspace}
\def\BRtaumtoe  {\ensuremath{\BR(\taum \to \en)}\xspace}
\def\BRtauptomu {\ensuremath{\BR(\taup \to \mup)}\xspace}
\def\BRtaumtomu {\ensuremath{\BR(\taum \to \mun)}\xspace}

\newcommand{\etaprepp}{\ensuremath{\etapr \to \eta \pipi}\xspace}
\newcommand{\etaprrg} {\ensuremath{\etapr \to \rho^0 \g}\xspace}

\def\bdpsikstar {\ensuremath{\Bd \to \jpsi \Kstarz}\xspace}
\def\bspsiphi   {\ensuremath{\Bs \to \jpsi \phi}\xspace}
\def\bsphimm    {\ensuremath{\Bs \to \phi\mup\mun}\xspace}
\def\bdkmm      {\ensuremath{\Bz \to K\mup\mun}\xspace}
\def\bpsiks     {\ensuremath{\Bz \to \jpsi \KS}\xspace}
\def\bpsikst    {\ensuremath{\Bz \to \jpsi \Kstar}\xspace}
\def\bpsikl     {\ensuremath{\Bz \to \jpsi \KL}\xspace}
\def\bpsikzeropi{\ensuremath{\Bz \to \jpsi \Kstarz (\to \KL \piz)}\xspace}
\def\bpsikpluspi{\ensuremath{\Bu \to \jpsi \Kstarp (\to \KL \pip)}\xspace}
\def\bpsikpi    {\ensuremath{\Bz/\Bzb \to \jpsi (\to \mumu) \Kpm \pimp}\xspace}
\def\bspsiphi   {\ensuremath{\Bs \to \jpsi \phi}\xspace}
\def\bupsik     {\ensuremath{\Bpm \to \jpsi \Kpm}\xspace}
\def\bupsimmk   {\ensuremath{\Bpm \to \jpsi (\to \mumu) \Kpm}\xspace}
\def\bpsiX      {\ensuremath{\Bz \to \jpsi \X}\xspace}

\def\Bzbtomu    {\ensuremath{\Bzb \to \mu \X}\xspace}
\def\Bzbtox     {\ensuremath{\Bzb \to \X}\xspace}
\def\Bztopipi   {\ensuremath{\Bz \to \pipi}\xspace}
\def\Bztokpi    {\ensuremath{\Bz \to \Kpm \pimp}\xspace}
\def\Bztorhopi  {\ensuremath{\Bz \to \rho^+ \pim}\xspace}
\def\Bztorhorho {\ensuremath{\Bz \to \rho \rho}\xspace}
\def\Bztokrho   {\ensuremath{\Bz \to K \rho}\xspace}
\def\Bztokstpi  {\ensuremath{\Bz \to \Kstar \pi}\xspace}
\def\Bztoapi    {\ensuremath{\Bz \to a_1 \pi}\xspace}
\def\Bztodd     {\ensuremath{\Bz \to \DpDm}\xspace}
\def\Bztodstd   {\ensuremath{\Bz \to \Dstarp \Dm}\xspace}
\def\Bztodstdst {\ensuremath{\Bz \to \Dstarp \Dstarm}\xspace}

\def\BtoDK      {\ensuremath{B \to DK}\xspace}
\def\Btodstlnu  {\ensuremath{B \to \Dstar \ell \nu}\xspace}
\def\Btodstdlnu {\ensuremath{B \to \Dstar(D) \ell \nu}\xspace}
\def\Btorholnu  {\ensuremath{B \to \rho \ell \nu}\xspace}
\def\Btopilnu   {\ensuremath{B \to \pi \ell \nu}\xspace}

\def\Btoetah    {\ensuremath{B \to \eta h}\xspace}
\def\Btoetaph   {\ensuremath{B \to \etapr h}\xspace}

\newcommand{\Betaprks}{\ensuremath{\Bz \to \etapr \KS}\xspace}
\newcommand{\Betaprkz}{\ensuremath{\Bz \to \etapr \Kz}\xspace}

\def\btosgam    {\ensuremath{b \to s \g}\xspace}
\def\btodgam    {\ensuremath{b \to d \g}\xspace}
\def\btosll     {\ensuremath{b \to s \ellell}\xspace}
\def\btosmm     {\ensuremath{b \to s \mup\mun}\xspace}
\def\btosnunu   {\ensuremath{b \to s \nunub}\xspace}
\def\btosgaga   {\ensuremath{b \to s \gaga}\xspace}
\def\btosglue   {\ensuremath{b \to s g}\xspace}

\def\bbmumuX    {\ensuremath{\bbbar\to \mu^+\mu^-+X}\xspace}
\def\ccmumuX    {\ensuremath{\ccbar\to \mu^+\mu^-+X}\xspace}
\def\bpsimmX    {\ensuremath{b\to \jpsi(\to\mup\mun) X}\xspace}

\def\upsbb   {\ensuremath{\FourS \to \BB}\xspace}
\def\upsbzbz {\ensuremath{\FourS \to \BzBzb}\xspace}
\def\upsbpbm {\ensuremath{\FourS \to \BpBm}\xspace}
\def\upspsikl{\ensuremath{\FourS \to (\bpsikl) (\Bzbtox)}\xspace}

\def\tauptoe    {\ensuremath{\taup \to \ep \nunub}\xspace}
\def\taumtoe    {\ensuremath{\taum \to \en \nunub}\xspace}
\def\tauptomu   {\ensuremath{\taup \to \mup \nunub}\xspace}
\def\taumtomu   {\ensuremath{\taum \to \mun \nunub}\xspace}
\def\tauptopi   {\ensuremath{\taup \to \pip \nub}\xspace}
\def\taumtopi   {\ensuremath{\taum \to \pim \nu}\xspace}

\def\ggtopi     {\ensuremath{\gaga \to \pipi}\xspace}
\def\ggtopiz    {\ensuremath{\gaga \to \ppz}\xspace}
\def\ggstox     {\ensuremath{\ggstar \to \X(1420) \to \kaon \kaon \pi}\xspace}
\def\ggstoeta   {\ensuremath{\ggstar \to \eta(550) \to \pipi \piz}\xspace}

\def\ptot       {\mbox{$p$}\xspace}
\def\pxy        {\mbox{$p_T$}\xspace}
\def\ptrel      {\mbox{$p_\perp^{\mathrm{rel}}$}\xspace}
\def\mes        {\mbox{$m_{\rm ES}$}\xspace}
\def\mec        {\mbox{$m_{\rm EC}$}\xspace}
\def\DeltaE     {\mbox{$\Delta E$}\xspace}

\def\pbcm {\ensuremath{p^*_{\Bz}}\xspace}

\def\mphi       {\mbox{$\phi$}\xspace}
\def\mtheta     {\mbox{$\theta$}\xspace}
\def\ctheta     {\mbox{$\cos\theta$}\xspace}

\newcommand{\eev}{\ensuremath{\mbox{\,Ee\kern -0.1em V}}\xspace}
\newcommand{\pev}{\ensuremath{\mbox{\,Pe\kern -0.1em V}}\xspace}
\newcommand{\tev}{\ensuremath{\mbox{\,Te\kern -0.1em V}}\xspace}
\renewcommand{\gev}{\ensuremath{\mbox{\,Ge\kern -0.1em V}}\xspace}
\newcommand{\mev}{\ensuremath{\mbox{\,Me\kern -0.1em V}}\xspace}
\newcommand{\kev}{\ensuremath{\mbox{\,ke\kern -0.1em V}}\xspace}
\newcommand{\ev} {\ensuremath{\mbox{\,e\kern -0.1em V}}\xspace}
\def\microEv         {\ensuremath{\,\mu\mbox{eV}}\xspace}  
\def\milliEv     {\ensuremath{\,\mbox{meV}}\xspace}  
\newcommand{\gevc}{\ensuremath{{\mbox{\,Ge\kern -0.1em V\!/}c}}\xspace}
\newcommand{\mevc}{\ensuremath{{\mbox{\,Me\kern -0.1em V\!/}c}}\xspace}
\newcommand{\gevcc}{\ensuremath{{\mbox{\,Ge\kern -0.1em V\!/}c^2}}\xspace}
\newcommand{\mevcc}{\ensuremath{{\mbox{\,Me\kern -0.1em V\!/}c^2}}\xspace}

\def\N   {\ensuremath{\mbox{\,N}\xspace}}

\def\syin {\ensuremath{^{\prime\prime}}\xspace}
\def\inch {\ensuremath{\rm \,in}\xspace} 
\def\ft   {\ensuremath{\rm \,ft}\xspace}
\def\km   {\ensuremath{\mbox{\,km}}\xspace}
\def\m    {\ensuremath{\mbox{\,m}}\xspace}
\def\cm   {\ensuremath{\mbox{\,cm}}\xspace}
\def\sr   {\ensuremath{\mbox{\,sr}}\xspace}
\def\cma  {\ensuremath{\mbox{\,cm}^2}\xspace}
\def\mm   {\ensuremath{\mbox{\,mm}\xspace}}
\def\mma  {\ensuremath{\mbox{\,mm}^2}\xspace}
\def\mum  {\ensuremath{\,\mu\mbox{m}\xspace}}
\def\muma {\ensuremath{\,\mu\mbox{m^2}}\xspace}
\def\fm   {\ensuremath{\mbox{\,fm}}\xspace}
\def\nm   {\ensuremath{\mbox{\,nm}}\xspace}   
\def\nb   {\ensuremath{\mbox{\,nb}}\xspace}
\def\barn      {\ensuremath{\mbox{\,b}}\xspace}
\def\mbarn     {\ensuremath{\mbox{\,mb}}\xspace}
\def\mb        {\ensuremath{\mbox{\,mb}}\xspace}
\def\pb        {\ensuremath{\mbox{\,pb}}\xspace}
\def\invmb     {\ensuremath{\mbox{\,mb}^{-1}}\xspace}
\def\invnb     {\ensuremath{\mbox{\,nb}^{-1}}\xspace}
\def\invpb     {\ensuremath{\mbox{\,pb}^{-1}}\xspace}
\def\ub        {\ensuremath{\,\mu\mbox{b}}\xspace}
\def\invub     {\ensuremath{\mbox{\,\ub}^{-1}}\xspace}
\def\fb        {\ensuremath{\mbox{\,fb}}\xspace}
\def\invfb     {\ensuremath{\mbox{\,fb}^{-1}}\xspace}
\def\ab        {\ensuremath{\mbox{\,ab}}\xspace}
\def\invab     {\ensuremath{\mbox{\,ab}^{-1}}\xspace}
\def\cms       {\ensuremath{\mbox{\,cm}^{-2}\mbox{s}^{-1}}\xspace}
\def\sqrts     {\ensuremath{\sqrt{s}}}

\def\mpc     {\ensuremath{\mbox{\,Mpc}}\xspace}

\def\kW   {\ensuremath{\mbox{\,kW}}\xspace}
\def\MW   {\ensuremath{\mbox{\,MW}}\xspace}
\def\mW   {\ensuremath{\mbox{\,mW}}\xspace}
\def\GW   {\ensuremath{\mbox{\,GW}}\xspace}

\def\Hz  {\ensuremath{\mbox{\, Hz}}\xspace}
\def\kHz {\ensuremath{\mbox{\, kHz}}\xspace}
\def\MHz {\ensuremath{\mbox{\, MHz}}\xspace}
\def\us   {\ensuremath{\,\mu\mbox{s}}\xspace}
\def\ns   {\ensuremath{\mbox{\,ns}}\xspace}
\def\ms   {\ensuremath{\mbox{\,ms}}\xspace}
\def\ps   {\ensuremath{\mbox{\,ps}}\xspace}
\def\fs   {\ensuremath{\mbox{\,fs}}\xspace}
\def\gm   {\ensuremath{\mbox{\,g}}\xspace}
\def\Gy{\ensuremath{\mbox{\,Gy}}\xspace}
\def\sec{\ensuremath{\mbox{\,s}}\xspace}       
\def\msec{\ensuremath{\mbox{\,ms}}\xspace}       
\def\usec{\ensuremath{\,\mu \mbox{s}}\xspace}       
\def\h          {\ensuremath{\mbox{\,h}\xspace}}

\def\kg         {\ensuremath{\mbox{\,kg}}\xspace}  
\def\gram       {\ensuremath{\mbox{\,g}}\xspace}  

\def\uTesla     {\ensuremath{\,\mu\mbox{T}}\xspace}  
\def\mTesla     {\ensuremath{\mbox{\,mT}}\xspace}  
\def\nTesla     {\ensuremath{\mbox{\,nT}}\xspace}  
\def\Tesla      {\ensuremath{\mbox{\,T}}\xspace}  
\def\Gauss      {\ensuremath{\mbox{\,G}}\xspace}  

\def\mA     {\ensuremath{\mbox{\,mA}}\xspace}  
\def\Ampere     {\ensuremath{\mbox{\,A}}\xspace}  
\def\Amp     {\ensuremath{\mbox{\,A}}\xspace}  
\def\Watt     {\ensuremath{\mbox{\,W}}\xspace}  

\def\Xrad {\ensuremath{X_0}\xspace}
\def\NIL{\ensuremath{\lambda_{int}}\xspace}
\let\dgr\degrees

\def\mbar        {\ensuremath{\mbox{\,mbar}}}   

\def\MVolts {\ensuremath{\mbox{\, MV}}\xspace}
\def\kVolts {\ensuremath{\mbox{\, kV}}\xspace}
\def\Volts {\ensuremath{\mbox{\, V}}\xspace}
\def\Volt  {\ensuremath{\mbox{\, V}}\xspace}
\def\atm   {\ensuremath{\mbox{\,atm}}\xspace}
\def\Ke    {\ensuremath{\mbox{\, K}}\xspace}
\def\mKe   {\ensuremath{\mbox{\, mK}}\xspace}
\def\mic  {\ensuremath{\,\mu{\rm C}}\xspace}
\def\krad {\ensuremath{\rm \,krad}\xspace}
\def\cmc  {\ensuremath{{\rm \,cm}^3}\xspace}
\def\yr   {\ensuremath{\rm \,yr}\xspace}
\def\hr   {\ensuremath{\rm \,hr}\xspace}
\def\degc {\ensuremath{^\circ}{C}\xspace}
\def\degk {\ensuremath{\mbox{K}}\xspace}
\def\degrees {\ensuremath{^{\circ}}\xspace}
\def\mrad {\ensuremath{\,\mbox{mrad}}\xspace}               
\def\urad {\ensuremath{\,\mu\mbox{rad}}\xspace}               
\def\rad{\ensuremath{\mbox{\,rad}}\xspace}
\def\mradhyph{\ensuremath{\rm -mr}\xspace}
\def\sx    {\ensuremath{\sigma_x}\xspace}
\def\sy    {\ensuremath{\sigma_y}\xspace}
\def\sz    {\ensuremath{\sigma_z}\xspace}
\def\order{{\ensuremath{\cal O}}\xspace}
\def\L{{\ensuremath{\cal L}}\xspace}
\def\calL{{\ensuremath{\cal L}}\xspace}
\def\calS{{\ensuremath{\cal S}}\xspace}
\def\calA{{\ensuremath{\cal A}}\xspace}
\def\calD{{\ensuremath{\cal D}}\xspace}
\def\calR{{\ensuremath{\cal R}}\xspace}
\def\ra                 {\ensuremath{\rightarrow}\xspace}
\def\to                 {\ensuremath{\rightarrow}\xspace}

\renewcommand{\stat}{\ensuremath{\mathrm{(stat)}}\xspace}
\renewcommand{\syst}{\ensuremath{\mathrm{(syst)}}\xspace}

\newcommand{\sstat}{\ensuremath{\sigma_{\mathrm{stat}}}\xspace}
\newcommand{\ssyst}{\ensuremath{\sigma_{\mathrm{syst}}}\xspace}

\def\pep2{PEP-II}
\def\BF{$B$ Factory}
\def\abf {asymmetric \BF}
\def\sx    {\ensuremath{\sigma_x}\xspace}
\def\sy    {\ensuremath{\sigma_y}\xspace}
\def\sz    {\ensuremath{\sigma_z}\xspace}

\newcommand{\inverse}{\ensuremath{^{-1}}\xspace}
\newcommand{\dedx}{\ensuremath{\mathrm{d}\hspace{-0.1em}E/\mathrm{d}x}\xspace}
\newcommand{\chisq}{\ensuremath{\chi^2}\xspace}
\newcommand{\delm}{\ensuremath{m_{\dstr}-m_{\dz}}\xspace}
\newcommand{\lum} {\ensuremath{\mathcal{L}}\xspace}

\def\gsim{{~\raise.15em\hbox{$>$}\kern-.85em
          \lower.35em\hbox{$\sim$}~}\xspace}
\def\lsim{{~\raise.15em\hbox{$<$}\kern-.85em
          \lower.35em\hbox{$\sim$}~}\xspace}

\def\qsq                {\ensuremath{q^2}\xspace}
\def\kbytes     {\ensuremath{{\rm \,kbytes}}\xspace}
\def\kbsps      {\ensuremath{{\rm \,kbytes/s}}\xspace}
\def\kbits      {\ensuremath{{\rm \,kbits}}\xspace}
\def\kbsps      {\ensuremath{{\rm \,kbits/s}}\xspace}
\def\mbsps      {\ensuremath{{\rm \,Mbits/s}}\xspace}
\def\mbytes     {\ensuremath{{\rm \,Mbytes}}\xspace}
\def\mbps       {\ensuremath{{\rm \,Mbyte/s}}\xspace}
\def\mbsps      {\ensuremath{{\rm \,Mbytes/s}}\xspace}
\def\gbsps      {\ensuremath{{\rm \,Gbits/s}}\xspace}
\def\gbytes     {\ensuremath{{\rm \,Gbytes}}\xspace}
\def\gbsps      {\ensuremath{{\rm \,Gbytes/s}}\xspace}
\def\tbytes     {\ensuremath{{\rm \,Tbytes}}\xspace}
\def\tbpy       {\ensuremath{{\rm \,Tbytes/yr}}\xspace}
\def\tb         {\ensuremath{\tan\beta}\xspace}

\newcommand{\as}{\ensuremath{\alpha_{\scriptscriptstyle S}}\xspace}
\newcommand{\asp}{\ensuremath{{\alpha_{\scriptscriptstyle S}\over\pi}}\xspace}
\newcommand{\MSb}{\ensuremath{\overline{\mathrm{MS}}}\xspace}
\newcommand{\LMSb}{%
  \ensuremath{\Lambda_{\overline{\scriptscriptstyle\mathrm{MS}}}}\xspace
}

\def\eps{\varepsilon\xspace}
\def\epsK{\varepsilon_K\xspace}
\def\epsB{\varepsilon_B\xspace}
\def\epsp{\varepsilon^\prime_K\xspace}

\def\CP                {\ensuremath{C\!P}\xspace}
\def\CPT               {\ensuremath{C\!PT}\xspace} 

\def\epstag  {\ensuremath{\varepsilon_{\rm tag}}\xspace}
\def\tagfac  {\ensuremath{\epstag(1-2w)^2}\xspace}

\def\rhobar {\ensuremath{\overline \rho}\xspace}
\def\etabar {\ensuremath{\overline \eta}\xspace}
\def\meas {\ensuremath{|V_{cb}|, |\frac{V_{ub}}{V_{cb}}|,
|\varepsilon_K|, \Delta m_{B_d}}\xspace}

\def\Vud  {\ensuremath{|V_{ud}|}\xspace}
\def\Vcd  {\ensuremath{|V_{cd}|}\xspace}
\def\Vtd  {\ensuremath{|V_{td}|}\xspace}
\def\Vus  {\ensuremath{|V_{us}|}\xspace}
\def\Vcs  {\ensuremath{|V_{cs}|}\xspace}
\def\Vts  {\ensuremath{|V_{ts}|}\xspace}
\def\Vtd  {\ensuremath{|V_{td}|}\xspace}
\def\Vub  {\ensuremath{|V_{ub}|}\xspace}
\def\Vcb  {\ensuremath{|V_{cb}|}\xspace}
\def\Vtb  {\ensuremath{|V_{tb}|}\xspace}
\def\stwoa{\ensuremath{\sin\! 2 \alpha  }\xspace}
\def\stwob{\ensuremath{\sin\! 2 \beta   }\xspace}
\def\stwog{\ensuremath{\sin\! 2 \gamma  }\xspace}
\def\mistag{\ensuremath{w}\xspace}
\def\dilution{\ensuremath{\cal D}\xspace}
\def\deltaz{\ensuremath{{\rm \Delta}z}\xspace}
\def\deltat{\ensuremath{{\rm \Delta}t}\xspace}
\def\deltamd{\ensuremath{{\rm \Delta}m_d}\xspace}

\newcommand{\fsubd}{\ensuremath{f_D}}\xspace
\newcommand{\fds}{\ensuremath{f_{D_s}}\xspace}
\newcommand{\fsubb}{\ensuremath{f_B}\xspace}
\newcommand{\fbd}{\ensuremath{f_{B_d}}\xspace}
\newcommand{\fbs}{\ensuremath{f_{B_s}}\xspace}
\newcommand{\bsubb}{\ensuremath{B_B}\xspace}
\newcommand{\bbd}{\ensuremath{B_{B_d}}\xspace}
\newcommand{\bbs}{\ensuremath{B_{B_s}}\xspace}
\newcommand{\rgbb}{\ensuremath{\hat{B}_B}\xspace}
\newcommand{\rgbbd}{\ensuremath{\hat{B}_{B_d}}\xspace}
\newcommand{\rgbbs}{\ensuremath{\hat{B}_{B_s}}\xspace}
\newcommand{\rgbk}{\ensuremath{\hat{B}_K}\xspace}
\newcommand{\lqcd}{\ensuremath{\Lambda_{\mathrm{QCD}}}\xspace}
\newcommand{\secref}[1]{Section~\ref{sec:#1}}
\newcommand{\subsecref}[1]{Section~\ref{subsec:#1}}
\newcommand{\figref}[1]{Figure~\ref{fig:#1}}
\newcommand{\tabref}[1]{Table~\ref{tab:#1}}
\newcommand{\epjBase}        {Eur.\ Phys.\ Jour.\xspace}
\newcommand{\jprlBase}       {Phys.\ Rev.\ Lett.\xspace}
\newcommand{\jprBase}        {Phys.\ Rev.\xspace}
\newcommand{\jplBase}        {Phys.\ Lett.\xspace}
\newcommand{\nimBaseA}       {Nucl.\ Instr.\ Meth.\xspace}
\newcommand{\nimBaseB}       {Nucl.\ Instr.\ and Meth.\xspace}
\newcommand{\nimBaseC}       {Nucl.\ Instr.\ and Methods\xspace}
\newcommand{\nimBaseD}       {Nucl.\ Instrum.\ Methods\xspace}
\newcommand{\npBase}         {Nucl.\ Phys.\xspace}
\newcommand{\zpBase}         {Z.\ Phys.\xspace}

\newcommand{\apas}      [1]  {{Acta Phys.\ Austr.\ Suppl.\ {\bf #1}}}
\newcommand{\app}       [1]  {{Acta Phys.\ Polon.\ {\bf #1}}}
\newcommand{\ace}       [1]  {{Adv.\ Cry.\ Eng.\ {\bf #1}}}
\newcommand{\anp}       [1]  {{Adv.\ Nucl.\ Phys.\ {\bf #1}}}
\newcommand{\annp}      [1]  {{Ann.\ Phys.\ {\bf #1}}}
\newcommand{\araa}      [1]  {{Ann.\ Rev.\ Astr.\ Ap.\ {\bf #1}}}
\newcommand{\arnps}     [1]  {{Ann.\ Rev.\ Nucl.\ Part.\ Sci.\ {\bf #1}}}
\newcommand{\arns}      [1]  {{Ann.\ Rev.\ Nucl.\ Sci.\ {\bf #1}}}
\newcommand{\appopt}    [1]  {{Appl.\ Opt.\ {\bf #1}}}
\newcommand{\japj}      [1]  {{Astro.\ Phys.\ J.\ {\bf #1}}}
\newcommand{\baps}      [1]  {{Bull.\ Am.\ Phys.\ Soc.\ {\bf #1}}}
\newcommand{\seis}      [1]  {{Bull.\ Seismological Soc.\ of Am.\ {\bf #1}}}
\newcommand{\cmp}       [1]  {{Commun.\ Math.\ Phys.\ {\bf #1}}}
\newcommand{\cnpp}      [1]  {{Comm.\ Nucl.\ Part.\ Phys.\ {\bf #1}}}
\newcommand{\cpc}       [1]  {{Comput.\ Phys.\ Commun.\ {\bf #1}}}
\newcommand{\epj}       [1]  {\epjBase\ {\bf #1}}
\newcommand{\epjc}      [1]  {\epjBase\ C~{\bf #1}}
\newcommand{\fizika}    [1]  {{Fizika~{\bf #1}}}
\newcommand{\fp}        [1]  {{Fortschr.\ Phys.\ {\bf #1}}}
\newcommand{\ited}      [1]  {{IEEE Trans.\ Electron.\ Devices~{\bf #1}}}
\newcommand{\itns}      [1]  {{IEEE Trans.\ Nucl.\ Sci.\ {\bf #1}}}
\newcommand{\ijqe}      [1]  {{IEEE J.\ Quantum Electron.\ {\bf #1}}}
\newcommand{\ijmp}      [1]  {{Int.\ Jour.\ Mod.\ Phys.\ {\bf #1}}}
\newcommand{\ijmpa}     [1]  {{Int.\ J.\ Mod.\ Phys.\ {\bf A{\bf #1}}}}
\newcommand{\jl}        [1]  {{JETP Lett.\ {\bf #1}}}
\newcommand{\jetp}      [1]  {{JETP~{\bf #1}}}
\newcommand{\jpg}       [1]  {{J.\ Phys.\ {\bf G{\bf #1}}}}
\newcommand{\jap}       [1]  {{J.\ Appl.\ Phys.\ {\bf #1}}}
\newcommand{\jmp}       [1]  {{J.\ Math.\ Phys.\ {\bf #1}}}
\newcommand{\jmes}      [1]  {{J.\ Micro.\ Elec.\ Sys.\ {\bf #1}}}
\newcommand{\mpl}       [1]  {{Mod.\ Phys.\ Lett.\ {\bf #1}}}
\newcommand{\nim}       [1]  {\nimBaseC~{\bf #1}}
\newcommand{\nima}      [1]  {\nimBaseC~A~{\bf #1}}
\newcommand{\np}        [1]  {\npBase\ {\bf #1}}
\newcommand{\npb}       [1]  {\npBase\ B~{\bf #1}}
\newcommand{\npps}      [1]  {{Nucl.\ Phys.\ Proc.\ Suppl.\ {\bf #1}}}
\newcommand{\npaps}     [1]  {{Nucl.\ Phys.\ A~Proc.\ Suppl.\ {\bf #1}}}
\newcommand{\npbps}     [1]  {{Nucl.\ Phys.\ B~Proc.\ Suppl.\ {\bf #1}}}

\newcommand{\ncim}      [1]  {{Nuo.\ Cim.\ {\bf #1}}}
\newcommand{\optl}      [1]  {{Opt.\ Lett.\ {\bf #1}}}
\newcommand{\optcom}    [1]  {{Opt.\ Commun.\ {\bf #1}}}
\newcommand{\partacc}   [1]  {{Particle Acclerators~{\bf #1}}}
\newcommand{\pan}       [1]  {{Phys.\ Atom.\ Nuclei~{\bf #1}}}
\newcommand{\pflu}      [1]  {{Physics of Fluids~{\bf #1}}}
\newcommand{\ptoday}    [1]  {{Physics Today~{\bf #1}}}
\newcommand{\jpl}       [1]  {\jplBase\ {\bf #1}}
\newcommand{\plb}       [1]  {\jplBase\ B~{\bf #1}}
\newcommand{\prep}      [1]  {{Phys.\ Rep.\ {\bf #1}}}
\newcommand{\jprl}      [1]  {\jprlBase\ {\bf #1}}
\newcommand{\pr}        [1]  {\jprBase\ {\bf #1}}
\newcommand{\jpra}      [1]  {\jprBase\ A~{\bf #1}}
\newcommand{\jprd}      [1]  {\jprBase\ D~{\bf #1}}
\newcommand{\jpre}      [1]  {\jprBase\ E~{\bf #1}}

\newcommand{\prsl}      [1]  {{Proc.\ Roy.\ Soc.\ Lond.\ {\bf #1}}}
\newcommand{\ppnp}      [1]  {{Prog.\ Part.\ Nucl.\ Phys.\ {\bf #1}}}
\newcommand{\progtp}    [1]  {{Prog.\ Th.\ Phys.\ {\bf #1}}}
\newcommand{\rpp}       [1]  {{Rep.\ Prog.\ Phys.\ {\bf #1}}}
\newcommand{\jrmp}      [1]  {{Rev.\ Mod.\ Phys.\ {\bf #1}}}  
\newcommand{\rsi}       [1]  {{Rev.\ Sci.\ Instr.\ {\bf #1}}}
\newcommand{\sci}       [1]  {{Science~{\bf #1}}}
\newcommand{\sjnp}      [1]  {{Sov.\ J.\ Nucl.\ Phys.\ {\bf #1}}}
\newcommand{\spd}       [1]  {{Sov.\ Phys.\ Dokl.\ {\bf #1}}}
\newcommand{\spu}       [1]  {{Sov.\ Phys.\ Usp.\ {\bf #1}}}
\newcommand{\tmf}       [1]  {{Teor.\ Mat.\ Fiz.\ {\bf #1}}}
\newcommand{\yf}        [1]  {{Yad.\ Fiz.\ {\bf #1}}}
\newcommand{\zpr}       [1]  {{ZhETF Pis.\ Red.\ {\bf #1}}}

\newcommand{\hepex}     [1]  {hep-ex/{#1}}
\newcommand{\hepph}     [1]  {hep-ph/{#1}}
\newcommand{\hepth}     [1]  {hep-th/{#1}}
\def\aslund     {\mbox{\tt Aslund}\xspace}
\def\bbsim      {\mbox{\tt BBsim}\xspace}
\def\beast      {\mbox{\tt Beast}\xspace}
\def\beget      {\mbox{\tt Beget}\xspace}
\def\Bta        {\mbox{\tt Beta}\xspace}
\def\betakfit   {\mbox{\tt BetaKfit}\xspace}
\def\cornelius  {\mbox{\tt Cornelius}\xspace}
\def\evtgen     {\mbox{\tt EvtGen}\xspace}
\def\euclid     {\mbox{\tt Euclid}\xspace}
\def\fitver     {\mbox{\tt FitVer}\xspace}
\def\fluka      {\mbox{\tt Fluka}\xspace}
\def\fortran    {\mbox{\tt Fortran}\xspace}
\def\gcalor     {\mbox{\tt GCalor}\xspace}
\def\geant      {\mbox{\tt GEANT}\xspace}
\def\gheisha    {\mbox{\tt Gheisha}\xspace}
\def\hemicosm   {\mbox{\tt HemiCosm}\xspace}
\def\hepevt     {\mbox{\tt{/HepEvt/}}\xspace}
\def\jetset74   {\mbox{\tt Jetset \hspace{-0.5em}7.\hspace{-0.2em}4}\xspace}
\def\koralb     {\mbox{\tt KoralB}\xspace}
\def\minuit     {\mbox{\tt Minuit}\xspace}
\def\objegs     {\mbox{\tt Objegs}\xspace}
\def\paw        {\mbox{\tt Paw}\xspace}
\def\root       {\mbox{\tt Root}\xspace}
\def\squaw      {\mbox{\tt Squaw}\xspace}
\def\stdhep     {\mbox{\tt StdHep}\xspace}
\def\trackerr   {\mbox{\tt TrackErr}\xspace}
\def\turtle     {\mbox{\tt Decay~Turtle}\xspace}

\def\pt{\ensuremath{{p_\mathrm{T}}}}
\def\ptmu{\ensuremath{{p^\mu_\mathrm{T}}}}
\def\etamu{\ensuremath{{\eta^\mu}}}

\def\vdef #1{\expandafter\def\csname #1\endcsname }
\def\vuse #1{\csname #1\endcsname}
\def\vu   #1{\csname #1\endcsname}

\vdef{n:result}{1.32\pm0.01 (\mathrm{stat}) \pm0.30 (\mathrm{syst}) \pm0.15 (\mathrm{lumi})}

\abstract{ A measurement of the b-hadron production cross section
  in proton-proton collisions at $\sqrt{s}=7\,\mathrm{TeV}$ is
  presented.  The dataset, corresponding to
  $85\,\mathrm{nb}^{-1}$, was recorded with the CMS experiment at the LHC
  using a low-threshold single-muon trigger.
  Events are selected by the presence
  of a muon with transverse momentum $p^\mu_\mathrm{T} > 6\,\mathrm{GeV}$
  with respect to the beam direction and pseudorapidity $|\eta^\mu| <
  2.1$. The transverse momentum of the muon with respect to the
  closest jet discriminates events containing b hadrons from background.
  The inclusive b-hadron production cross section is presented as a function of
  muon transverse momentum and pseudorapidity.
  The measured total cross section in the kinematic acceptance is $\sigma(\text{pp}\rightarrow \text{b}+X\rightarrow \mu + X')=1.32\pm0.01 (\mathrm{stat}) \pm0.30 (\mathrm{syst}) \pm0.15 (\mathrm{lumi}) \ub$.}
\hypersetup{%
pdfauthor={CMS Collaboration},%
pdftitle={Inclusive b-hadron production cross section with muons in pp collisions at sqrt(s) = 7 TeV},%
pdfsubject={CMS},%
pdfkeywords={CMS, physics}}

\maketitle 

\vdef{n:nevents} {157\,783}   
\vdef{n:lumi} {85}   
\vdef{n:lumiError} {9}   

\vdef{n:pythia}{1.8}
\vdef{n:mcnlo} {0.84^{+0.36}_{-0.19}({\mathrm{scale}}) \pm0.08({m_\text{b}}) \pm 0.04({\mathrm{pdf}})}

\vdef{n:ptmin} {6}  
\section{Introduction}
\label{s:introduction}
Measurements of b-hadron production in proton-proton (pp) collisions at the Large Hadron Collider (LHC)
are important tests of Quantum Chromodynamics (QCD)
in a new kinematical region.
Results on b-hadron production in proton-antiproton collisions at the lower center-of-mass energies, $\sqrt{s}$, of the
CERN S$\bar{\text{p}}$pS Collider~\cite{UA1-87} 
and the Tevatron~\cite{Abe:1993sj,Abachi:1994kj,Abbott:1999se,Aaltonen:2009xn}
have aroused substantial interest because of tensions
between the experimental results and the theoretical
expectations~\cite{Cacciari:2002pa,Cacciari:2003uh}.
First results at the LHC from pp collisions at $\sqrt{s}=7 \tev$ have been reported by the LHCb collaboration for the forward rapidity region using semi-inclusive decays~\cite{Bediaga:2010gn} and by the CMS collaboration in the central rapidity region using fully reconstructed $B^+$ hadron decays~\cite{BPH-10-004}.

The b-quark production cross section in
hadron collisions has been computed at next-to-leading order (NLO) in
perturbative QCD~\cite{Nason:1987xz,Nason:1989zy,Beenakker:1990maa}.
The observed large scale dependence of the NLO results is considered
to be a symptom of large contributions from higher orders: small-$x$
effects~\cite{Collins:1991ty,Catani:1990eg}, where $x \sim
m_\text{b}/\sqrts$, are possibly relevant in the low-\pt\ domain, while multiple-gluon
radiation leads to large logarithms of $\pt/m_\text{b}$ and may be
important at high \pt~\cite{Cacciari:1993mq}. The resummed logarithms of
$\pt/m_\text{b}$ at next-to-leading-logarithmic accuracy have been matched to
the fixed-order NLO calculation for massive
quarks~\cite{Cacciari:1998it}.  At the non-perturbative level, the b-hadron
\pt\ spectrum depends strongly on the parametrization of the
fragmentation function~\cite{Frixione:1997ma}.  The b-quark
production cross section has also been studied in the general-mass
variable-flavor-number scheme~\cite{Kniehl:2008zza} and the $k_\mathrm{T}$
factorization QCD approach~\cite{Ryskin:2000bz,Jung:2001rp}.

In this paper we present an inclusive measurement of the production of b hadrons decaying into muons and jets
based on $85\,\mathrm{nb}^{-1}$ of data recorded by the CMS experiment using a low-threshold single-muon trigger.
Muons from b-hadron decays are distinguished from backgrounds based on their transverse momentum relative
to a nearby jet ($\ptrel$).

In Section~\ref{s:cmsdetector} a brief overview of
the CMS detector is given. Section~\ref{s:mcsimulation} discusses the
Monte Carlo (MC) simulation used.  Section~\ref{s:selection} describes
the event selection and analysis methodology.  The systematic errors
are addressed in Section~\ref{s:systematics} and the results are
presented in Section~\ref{s:results}.

\section{The CMS Detector }
\label{s:cmsdetector}

A detailed description of the CMS detector can be found in
Ref.~\cite{Adolphi:2008zzk}.  The subdetectors used for the present
analysis are the inner tracker, consisting of silicon pixel and
silicon strip layers, and the muon detectors.  The inner tracker is
immersed in a $3.8$\,T axial magnetic field.  The pixel tracker
consists of three barrel layers and two endcap disks at each barrel
end. The strip tracker has 10 barrel layers and 12 end-cap disks at
each barrel end.  Muons are measured in gas-ionization detectors
embedded in the steel return yokes. In the barrel, there is a drift
tube system interspersed with resistive plate chambers (RPCs), and in
the end-caps there is a cathode strip chamber system, also
interspersed with RPCs. The first-level (L1) trigger used in this
analysis is based on the muon system alone, while the high-level
trigger (HLT) uses additional information from the inner tracker.

The CMS experiment uses a right-handed coordinate system, with the
origin at the nominal LHC beam collision point, the $x$ axis pointing
towards the center of the LHC ring and the $z$ axis pointing along the
counterclockwise beam direction.  The polar angle $\theta$ is measured
from the positive $z$ axis and the pseudorapidity is defined by
$\eta=-\ln\,\tan (\theta/2)$.  The azimuthal angle $\phi$ is
measured from the positive $x$ axis in the plane perpendicular to the
beam.

\section{Monte Carlo Simulation}
\label{s:mcsimulation}

The MC event generator \PYTHIA~6.422~\cite{Sjostrand:2006za} is
used (with MSEL=1) to compute efficiencies and kinematic distributions.
\PYTHIA and \MCATNLO~3.4~\cite{Frixione:2002ik,Frixione:2003ei} predictions are
compared with the experimental results. 
The programs were run with their default
parameter settings, except when mentioned otherwise. 
The \PYTHIA event sample was simulated with
the CTEQ6L1~\cite{Pumplin:2002vw} PDF, a b-quark mass $m_\text{b} = 4.8\gev$,
and Peterson \etal\ fragmentation functions~\cite{Peterson:1982ak} for c and  quarks with 
parameters $\varepsilon_\text{c}=0.05$ and $\varepsilon_{b}=0.005$. 
The underlying event is simulated with the D6T tune~\cite{Fano:2007zz}.
Pileup events were not included in the simulation and play a negligible
role in the data sample used for this measurement.  
 
For comparison, additional event samples were generated where the
\EVTGEN~\cite{Lange:2001uf} program was used to decay the b hadrons.
Events generated by the \PYTHIA program were passed through a detailed
MC simulation of the CMS detector response based on
\GEANTfour~\cite{Agostinelli:2002hh}.
The \MCATNLO package has a NLO matrix element calculation interfaced to
the parton shower algorithms of the \HERWIG~\cite{Corcella:2000bw} package. 
A b quark mass of  $m_\text{b} = 4.75\gev$ and the 
CTEQ6M PDF set~\cite{Pumplin:2002vw} were used.
The events generated with \MCATNLO are studied only at the generator
level and are not passed through the detailed detector simulation.

\section{Data Selection and Analysis}
\label{s:selection}

This analysis is based on data collected in 2010 when the collider and
detector were fully operational and fulfilled the following
requirements: (1) Stable beam conditions, (2) stable magnetic field
inside CMS at the nominal value, (3) operational L1 and HLT, and (4) inner
tracker and muon stations at their nominal high-voltage settings.  The
data sample used in this analysis corresponds to an integrated
luminosity of $\clu=\vu{n:lumi} \pm \vu{n:lumiError} \invnb$~\cite{PAS-EWK-10-004}.

The events of interest are selected by a very loose single-muon trigger path.
The L1 muon trigger makes no explicit requirement on the muon momentum
transverse to the $z$ axis, $\pt$, although muons with $\pt < 3\GeV$
do not have sufficient momentum to be reconstructed in the barrel
region of the muon system.

In the HLT, a standalone muon reconstruction (with information from
the muon detectors only) is seeded by the parameters of the L1 muon
candidate. If the standalone muon candidate has $\pt > 3\gev$ it serves as a
seed in the global muon reconstruction, where a track in the inner
tracker is linked to the standalone muon, and further selection
requirements are applied on the transverse momentum
($\pt>3\gev$) and the impact parameter with respect to the beam spot
in the transverse plane ($|d_0|<2\cm$).

The offline event selection requires a reconstructed primary vertex with more than three tracks
and at least one muon candidate with $\pt >\vu{n:ptmin} \gev$ and pseudorapidity $\left | \eta \right |<2.1$
that fulfills a tight muon selection similar to that in Ref.~\cite{Khachatryan:2010xn}.
The muon candidates are required to be reconstructed by two independent algorithms,
one starting from segments in the muon chambers and one starting from inner-tracker information.
The inner track must be measured with at least 10 hits in the inner tracker, two of which must be
on pixel layers.
The inner-track fit and the global muon fit (including all inner tracker and muon detector hits)
are required to have a $\chi^2$ of less than 10 per degree of freedom and at least two muon segments
matching the inner track must be found.
Only muon candidates with transverse impact parameter with respect to the primary vertex $|d_0| < 2 \mm$ and
longitudinal impact parameter with respect to the primary vertex $|d_z| < 1\cm$ are accepted.

In events passing the trigger and event selections, all tracks
including the muon are clustered into track-jets~\cite{PAS-JME-10-006}
by the anti-\kt\ jet algorithm~\cite{Cacciari:2008gp} with
$R=0.5$. The tracks are selected with the following requirements: $0.3
< \pt < 500\gev$, $|z_0| < 2\cm$, and hits in at least 2 (5) layers of
the pixel (pixel and strip) detector. Only jets containing a muon are
accepted as b-jet candidates.

The jet direction and jet energy $E$ are calculated by summing the four-momenta of
all tracks in the jet except the muon. The pion mass hypothesis is assumed
for calculating the energy associated with a track.
The jet is required to contain at least one track
and to have a transverse energy $E_T=E \sin \theta_{\mathrm{jet}}$ of at least 1 GeV,
where $\theta_{\mathrm{jet}}$ is the polar angle of the jet direction.

The efficiency for identifying b jets is determined
in MC simulation for events in which the muon from a b-hadron decay
falls into the kinematic region of this measurement.
The efficiency for finding a jet containing the muon rises with
the muon \pt\, from 74\% at $6 \gev$ to almost 100\% for events containing a muon with $\pt >20\gev$.
The fraction of events in which the reconstructed jet containing the muon
is not matched to the b jet at the generator level is smaller than 7\% in
the lowest muon transverse momentum bin and asymptotically reaches a value of 2\% at large \pt.

From the momenta of the selected muon $(\vec{p}_{\mu})$ and the
associated track-jet $(\vec{p}_{j})$, the relative transverse momentum
of the muon with respect to its track-jet is calculated as
$\ptrel = |\vec{p}_\mu\times \vec{p}_{j}|/|\vec{p}_{j}|$.

\begin{figure}[!htb]
 \begin{center}
    \centerline{{\includegraphics[width=0.56\textwidth,angle=90]{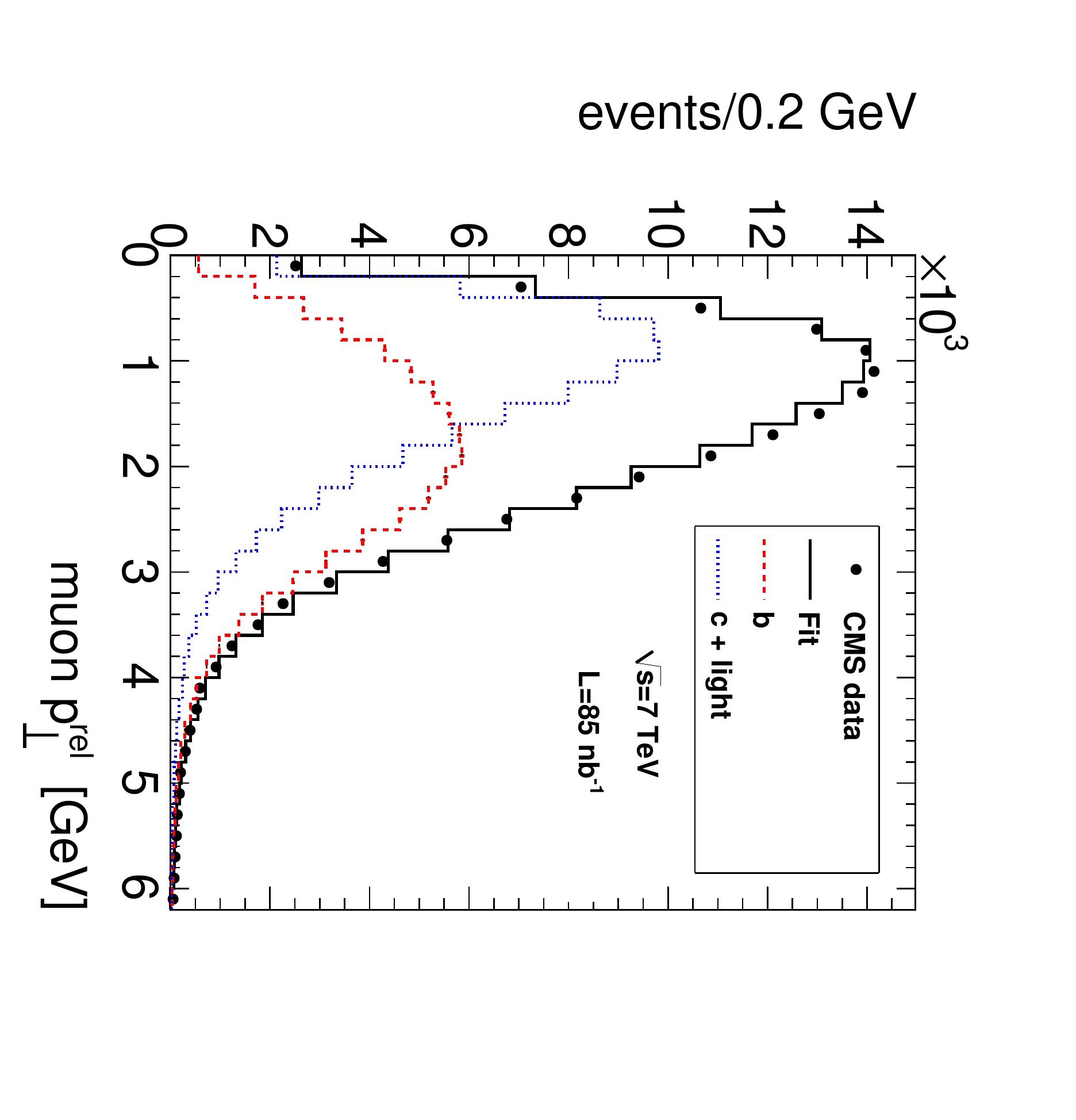}}}
  \caption{Distribution of muon transverse momentum \ptrel\ with respect to the closest
    track-jet in data and results of the maximum
    likelihood fit.   The black full circles
    correspond to the data distribution, while the black line is the
    result of the fitting procedure.
    The red dashed and the blue dotted line are the
    simulated b and cudsg distributions, respectively.
    \label{f:ptrel}}
 \end{center}
\end{figure}

A total of \vu{n:nevents} ~data events pass the selection. If an
event contains more than one muon of either charge, only the muon
with the largest transverse momentum \ptmu\ is kept. This affects
0.5\% of all data events.
\subsection{Fitting Procedure}

A fit to the observed \ptrel\ spectrum, based on distributions obtained
from simulation (signal and c$\bar{\text{c}}$) and data (the
remaining background), is used to determine the fraction of signal
events among all events passing the event selection. A binned
log-likelihood fit is performed, which takes into account the finite size of the
MC simulated sample~\cite{Barlow:1993dm}.

The distributions used in the fitting algorithm are determined separately
for the full sample and for each bin in muon transverse momentum and
pseudorapidity. Since the shape of the \ptrel~distribution in c and
light-quark/gluon (udsg) events cannot be distinguished by the fit,
the two background components are combined and a fit discriminating
the signal component against a single background component is
implemented. The udsg background is dominated by
hadrons misidentified as muons (mainly in-flight decays)
and is determined in data. Hadrons satisfying all muon track
selection criteria (without muon detector requirements) are weighted by
the misidentification probability and used instead of muons to determine
\ptrel. The  misidentification probability has been measured in
data~\cite{PAS-MUO-10-002}. The c background  is determined from MC
simulation.
Muons from sources other than b, c and udsg events are neglected.
The largest contribution to the muon event sample from these sources
is expected in the highest \ptmu\ bin (3\%, from $W$ decays).

The result of the fit in the full sample is displayed in
Fig.~\ref{f:ptrel}.  Extensive tests to validate the fitting
procedure were performed~\cite{LeaDiss} with repeated fits of MC
pseudo-experiments obtained by appropriate random variations. A
satisfactory performance of the fit was observed: the fit result does
not show a significant bias and the errors are properly calculated by
the fitter.  The stability of the fit was proven by repeated fits with varied binning.
The signal fractions have also been determined with particle flow
jets~\cite{CMS-PAS-PFT-10-002} and with a fit to the muon impact
parameter distribution.  The results are consistent with the
fit using track-jets within the systematic uncertainty.

\section{Systematic Uncertainties}
\label{s:systematics}

The systematic uncertainties of this analysis are dominated by the shapes of
the \ptrel\ distributions used in the fitting procedure.

\begin{table}[!htb]
 \begin{center}
  \caption{Summary of systematic cross section uncertainties. The systematic uncertainty
    can vary depending on the muon transverse momentum and
    pseudorapidity as indicated by the range.}

  \begin{tabular}{l r}
  \\
   \hline
   source   &  cross section uncertainty (\%)\\
   \hline
   Trigger efficiency                    &     5   \\
   Muon reconstruction efficiency        &     3   \\
   Hadron tracking efficiency            &     2   \\
   b \ptrel shape uncertainty          &  $\le$ 21   \\
   Background \ptrel shape uncertainty & 2--14   \\
   Background composition                &  3--6   \\
   Production mechanism                  &  2--5   \\
   Fragmentation                         &  1--4   \\
   Decay                                 &     3   \\
   Underlying event                      &    10   \\
   Luminosity                            &    11   \\
   \hline
  \end{tabular}
  \label{t:systematics}
 \end{center}
\end{table}

The signal \ptrel\ distribution is validated with data through a control sample
enriched in b decays.  Selecting muons with a large impact parameter
significance of $|d_0|/\sigma_{d_0} > 12$, where $d_0$ is the uncertainty
of the impact parameter measurement, results in an event sample with an expected b fraction
of about 85\%.
Small adjustments of the shape of the distributions by rescaling \ptrel\ improve the
agreement between data and simulation in the b-enriched region and in the full sample.
They result in variations of the measured cross section of up to 21\% that are taken as a systematic uncertainty.

The background consists of contributions from c$\bar{\text{c}}$ events and
from light-quark and gluon events, where a hadron is misidentified
as a muon.
Both contributions are similar in shape and magnitude.
The c fraction of the background is expected to rise with increasing muon \pt.
The fit does not separately determine the c and udsg
content of the sample. Two effects can introduce a systematic
uncertainty. (1) The udsg distribution determined from data could be biased.
Using the \PYTHIA-derived udsg background introduces a difference to
the reference fit of 2--14\%, depending on the muon transverse
momentum and pseudorapidity bin.  (2) If the c fraction of the
non-b background in the data was different from the value used in
combining the backgrounds, the fitted b fraction could change. The MC-simulation
predicts a c fraction of 50--70\% in the non-b
background depending on the muon transverse momentum. This fraction
depends on the modeling of charm semileptonic decays and on the hadron
misidentificaton probability. Varying the c vs.~udsg fraction by $\pm20\%$
leads to a systematic uncertainty of 3--6\%.

The muon trigger efficiency has been determined
from data with an uncertainty of 5\% using independent triggers.
The muon reconstruction efficiency is known to a precision of 3\%.
The tracking efficiency for hadrons is known with a precision of
about 4\%~\cite{PAS-TRK-10-002}, which induces a systematic uncertainty of
2\% on the number of events passing the event selection.

In \PYTHIA, the production of a b$\bar{\text{b}}$ pair can be separated into flavor
creation (19\% of the selected events), flavor excitation (56\%),
and gluon splitting (25\%). The event selection efficiencies are
71\%, 72\%, and 76\%, respectively.  Reweighting the events from the
different production processes to reflect the difference between
\PYTHIA and \HERWIG leads to a systematic uncertainty of 2--5\%, depending
on the muon transverse momentum.
The uncertainty of the b quark fragmentation is studied by varying the parameter $\varepsilon_\text{b}$
between 0.003 and 0.010, which results in a systematic uncertainty of
1--4\% on the reconstruction efficiency. A sample generated with
\EVTGEN is used to investigate the uncertainty in modeling the
b-hadron decay properties. A systematic uncertainty of 3\% is
found. Varying the fraction of prompt $\text{b}\to \mu$ decays with respect
to $\text{b} \to \text{c} \to \mu$ decays within its
uncertainty~\cite{Nakamura:2010zzi} changes the measured cross section
by 1\%. Neither the muon trigger efficiency nor the track-jet
finding is affected significantly by the variation of the
fragmentation and decay parameters.
The track-jet reconstruction can be affected by the underlying event.
Using simulated event samples with different MC tunes
(D6T~\cite{Fano:2007zz}, Pro-Q20~\cite{Buckley:2009bj}, and CW~\cite{Khachatryan:2010pv})
for the efficiency and acceptance calculation changes the cross section of the order of 10 \%.
At the present stage of the CMS experiment, the integrated luminosity
recorded is known with an accuracy of 11\%~\cite{PAS-EWK-10-004}.

Table~\ref{t:systematics} summarizes the systematic uncertainties.

\section{Results}
\label{s:results}

The inclusive production cross section for b quarks decaying into muons is calculated as
\begin{displaymath}
 \sigma \equiv\sigma(\text{pp}\to \text{b} + X \to \mu +X')
 =\frac{N_\text{b}}{\mathcal L\, \varepsilon}\,,
\end{displaymath}
where $N_\text{b}$ is the number of selected b events in data.
No distinction is made between positive and negative muons; $N_\text{b}$ includes the process $\text{pp}\to \bar{\text{b}} + X \to \mu +X'$.
The efficiency $\varepsilon$ includes the trigger efficiency, $(88\pm5)$\%, the muon reconstruction efficiency, $(94\pm 3)$\%, and the
efficiency for associating a track-jet to the reconstructed muon,
$(77\pm8)$\%.

The result of the inclusive production cross section for b quarks decaying into muons within
the kinematic range $\ptmu> \vu{n:ptmin} \gev$ and  $|\etamu|<2.1$ is

\begin{displaymath}
 \sigma= \vu{n:result}\ub,
\end{displaymath}
where the first uncertainty is statistical, the second is systematic, and the third is associated with the estimation of the integrated luminosity.
For comparison, the inclusive b-quark production cross section predicted by \MCATNLO is
\begin{displaymath}
\sigma_{\rm MC@NLO}= \vu{n:mcnlo}  \ub,
\end{displaymath}

where the first uncertainty is due to variations in the QCD scale, the second to the b-quark mass, and the third to the
parton distribution function.
The value of the scale uncertainty is obtained by varying the QCD renormalization
and factorization scales
as described in Ref.~\cite{Cacciari:2003uh}.
The b-quark mass was varied between $4.5 \GeV$ and $5.0 \GeV$ and
the uncertainty induced by the parton distribution function was evaluated using the eigenvector sets as described in Ref.~\cite{Pumplin:2002vw}.
The \PYTHIA prediction using the parameters described in Section~\ref{s:mcsimulation} is \vu{n:pythia} \ub.

The differential cross section is calculated from

\begin{displaymath}
 \frac{d\sigma(\text{pp}\to \text{b}+X\to \mu +X')}{dx}\bigg|_{\mathrm{bin\ } i}
 =\frac{N^{i}_\text{b}}{\mathcal L \,\varepsilon^i\, \Delta x^i}\,,
\end{displaymath}

where $x$ stands for the muon transverse momentum or the muon
pseudorapidity, and $\Delta x^i$ denotes the width of bin $i$.  The
number $N^{i}_\text{b}$ of selected b events in data and
the efficiency $\varepsilon_i$ are determined separately for each bin.

\begin{figure}[!htb]
 \begin{center}
   \includegraphics[angle=90,width=0.49\textwidth]{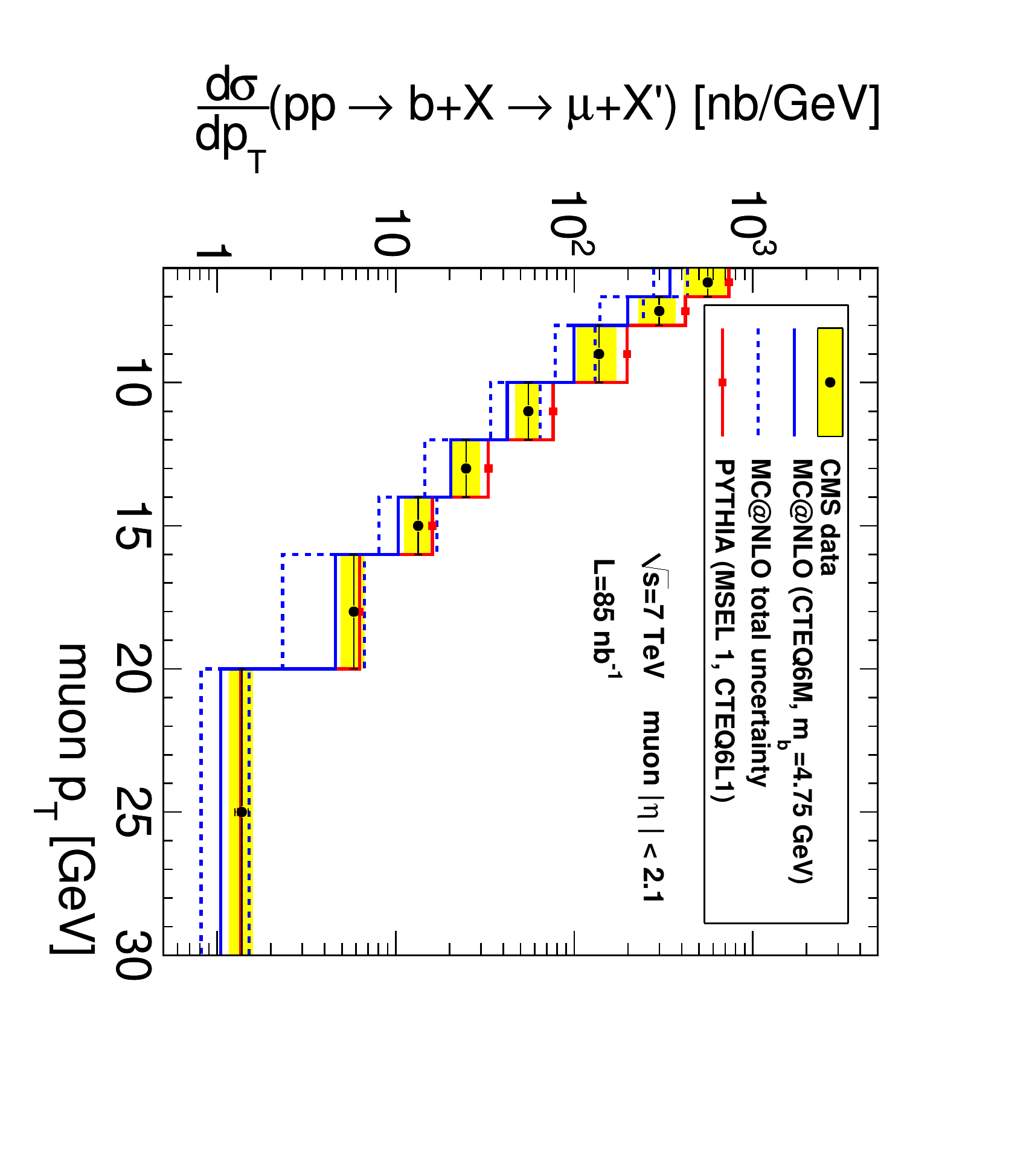}
   \includegraphics[angle=90,width=0.49\textwidth]{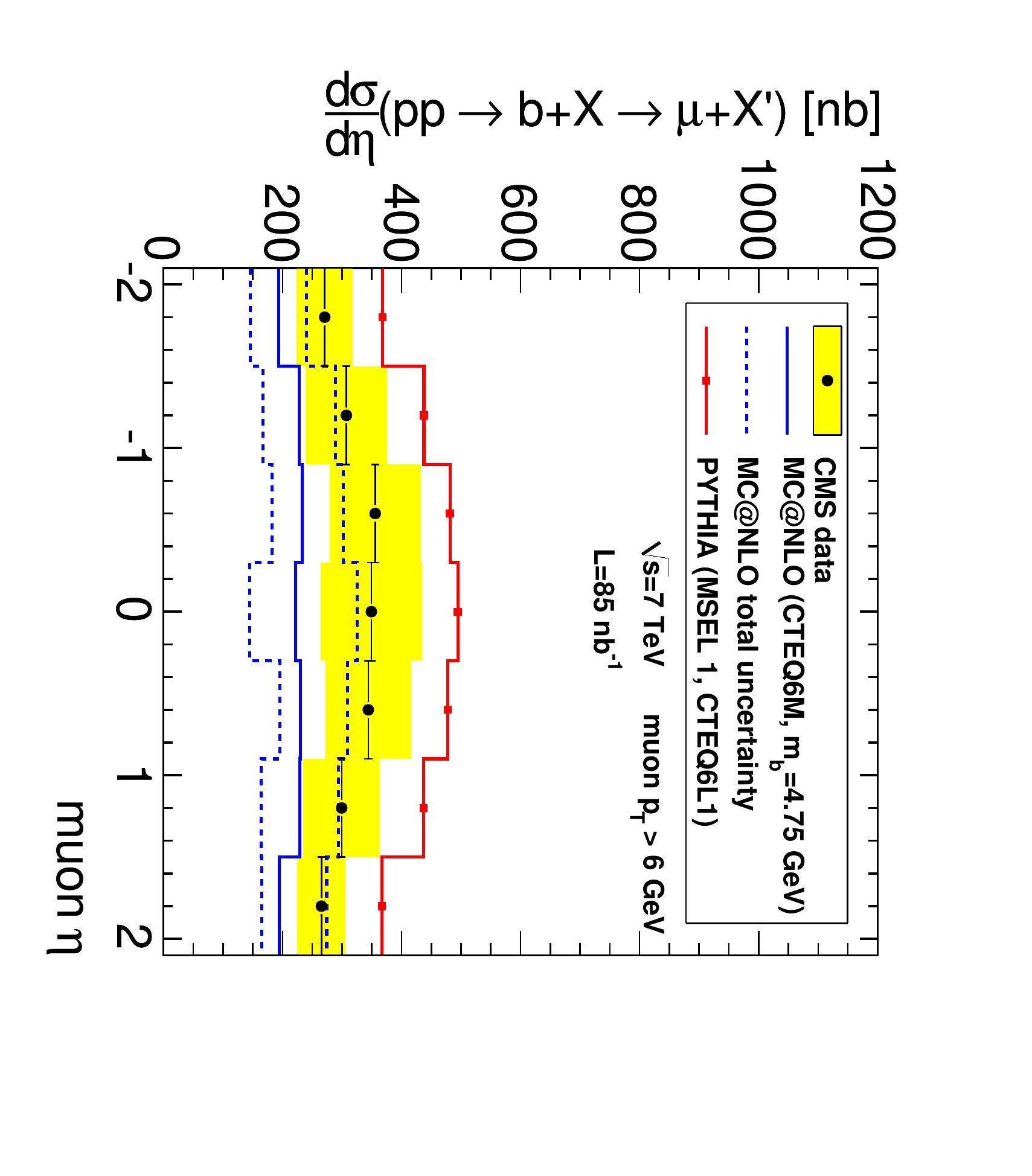}
  \caption{Differential cross section (left)
    $\frac{d\sigma}{d\ptmu}(\text{pp}\to \text{b}+X\to \mu +X', |\etamu|<2.1)$, and
    (right) $\frac{d\sigma}{d\etamu}(\text{pp}\to \text{b}+X\to \mu +X', \ptmu>
    \vu{n:ptmin} \gev)$. The two possible muon charges are not distinguished and the process $\text{pp}\to \bar{\text{b}}+X\to \mu +X'$ is included.
    The black points are the CMS measurements. Vertical error bars showing the statistical error are smaller than the point size in most bins, the horizontal bars indicate the bin width. The yellow band shows the quadratic sum of statistical and systematic
    uncertainties. The systematic uncertainty (11\%) of the luminosity
    measurement is not included.
    The solid blue line shows the \MCATNLO result and the dashed blue lines illustrate the
    theoretical uncertainty as described in the text.
    The solid red line with dots shows the \PYTHIA result.
    \label{f:dspt}}
 \end{center}
\end{figure}

The results of the differential b-quark production cross section as a
function of the muon transverse momentum and pseudorapidity are shown
in Fig.~\ref{f:dspt} and summarized in Table~\ref{t:respteta}.  The
data lie between the \PYTHIA and the \MCATNLO predictions. The
observed shapes of the kinematic distributions are described
reasonably well by both programs.
The integral of the differential cross section is consistent with the
cross section determined from the full sample.

\begin{table}[!htb]
 \begin{center}
  \caption{Differential cross sections
    ${d\sigma}/{d\ptmu}$ for $|\etamu|<2.1$ in bins of muon transverse
    momentum and  ${d\sigma}/{d\etamu}$  for $\ptmu>\vu{n:ptmin} \gev$ in bins of muon pseudorapidity.
    The number of b events ($N_\text{b}$, including $\bar{\text{b}}$ events) determined by the fit,
    the efficiency ($\varepsilon$) of the online and offline event
    selection, and the differential cross section, together with its
    relative statistical and systematic uncertainties, are
    given. A common uncertainty on the luminosity of 11\% is not included. }
  \label{t:respteta}

  \begin{tabular}{l c c c c c }
    \\
    \hline
    $\ptmu$ [GeV]&  $N_\text{b}$ &  $\varepsilon$ & $d\sigma/dp_\text{T}$ [nb/GeV] & stat (\%) & syst (\%)\\
    \hline
    6--7        & $26351\pm 523$  & $0.55\pm 0.01$  &    559    &  2 & 27 \\
    7--8        & $16016\pm 359$  & $0.63\pm 0.01$  &    299    &  2 & 23 \\
    8--10        & $16459\pm 332$  & $0.70\pm 0.01$  &    138    &  2 & 21 \\
    10--12        & $\phantom{1}7136\pm 209$  & $0.76\pm 0.02$  &     55    &  3 & 15 \\
    12--14      & $\phantom{1}3330\pm 146$  & $0.79\pm 0.02$  &     25    &  4 & 19 \\
    14--16        & $\phantom{1}1871\pm 102$  & $0.82\pm 0.04$  &     13    &  5 & 15 \\
    16--20        & $\phantom{1}1685\pm 99$  & $0.85\pm 0.04$  &    5.8    &  6 & 14 \\
    20--30       & $\phantom{11}969\pm 82\phantom{1}$  & $0.83\pm 0.04$  &    1.4    &  8 & 13 \\
    \hline
    $\eta^{\mu}$ &  $N_\text{b}$ &  $\varepsilon$  & $d\sigma/d\eta$ [nb] & stat & syst \\
    \hline
    (-2.1,-1.5)   & $\phantom{1}8452\pm 262$  & $0.61\pm 0.02$  &    271    &  3 & 18 \\
    (-1.5,-0.9)   & $9843\pm 276$  & $0.63\pm 0.02$  &    307    &  3 & 23 \\
    (-0.9,-0.3)   & $12476\pm 321$  & $0.68\pm 0.02$  &    356    &  3 & 23 \\
    (-0.3, 0.3)   & $11508\pm 315$  & $0.64\pm 0.02$  &    349    &  3 & 27 \\
    ( 0.3, 0.9)   & $11918\pm 312$  & $0.68\pm 0.02$  &    344    &  3 & 23 \\
    ( 0.9, 1.5)   & $\phantom{1}9330\pm 272$  & $0.61\pm 0.02$  &    299    &  3 & 24 \\
    ( 1.5, 2.1)   & $\phantom{1}8397\pm 255$  & $0.62\pm 0.02$  &    265    &  3 & 17 \\
    \hline
  \end{tabular}

 \end{center}
\end{table}

\section {Conclusions}
A measurement of the inclusive b-hadron production cross
section in the central rapidity region in proton-proton collisions 
at $\sqrt{s}=7\tev$ has been performed.
The measurement is based on a data sample corresponding to an integrated
luminosity of $\vu{n:lumi} \invnb$ recorded by the
CMS experiment during the first months of data taking in 2010
with a low-threshold single-muon trigger.

The result for the total inclusive production
cross section of b hadrons decaying into muons within the visible kinematic range is

\begin{displaymath}
 \sigma(\text{pp}\rightarrow \text{b}+X\rightarrow \mu + X')= \vu{n:result} \ub,
\end{displaymath}

where $\ptmu > \vu{n:ptmin} \gev, |\eta^\mu|<2.1$. 
The measured cross section is approximately 1.6 times higher than the \MCATNLO prediction,
but the difference is less than the theoretical and experimental uncertainties.
Differential cross sections have been measured as a function of muon transverse momentum and pseudorapidity. 
The observed shapes are reasonably well described by \MCATNLO.
 A similar pattern was recently found by this collaboration in the measurement of b production using fully reconstructed $\text{B}^+$ meson decays ~\cite{BPH-10-004}.

\section*{Acknowledgements}
We wish to congratulate our colleagues in the CERN accelerator departments for the excellent performance of the LHC machine. We thank the technical and administrative staff at CERN and other CMS institutes, and acknowledge support from: FMSR (Austria); FNRS and FWO (Belgium); CNPq, CAPES, FAPERJ, and FAPESP (Brazil); MES (Bulgaria); CERN; CAS, MoST, and NSFC (China); COLCIENCIAS (Colombia); MSES (Croatia); RPF (Cyprus); Academy of Sciences and NICPB (Estonia); Academy of Finland, ME, and HIP (Finland); CEA and CNRS/IN2P3 (France); BMBF, DFG, and HGF (Germany); GSRT (Greece); OTKA and NKTH (Hungary); DAE and DST (India); IPM (Iran); SFI (Ireland); INFN (Italy); NRF and WCU (Korea); LAS (Lithuania); CINVESTAV, CONACYT, SEP, and UASLP-FAI (Mexico); PAEC (Pakistan); SCSR (Poland); FCT (Portugal); JINR (Armenia, Belarus, Georgia, Ukraine, Uzbekistan); MST and MAE (Russia); MSTD (Serbia); MICINN and CPAN (Spain); Swiss Funding Agencies (Switzerland); NSC (Taipei); TUBITAK and TAEK (Turkey); STFC (United Kingdom); DOE and NSF (USA).

\bibliography{auto_generated}   

\cleardoublepage\appendix\section{The CMS Collaboration \label{app:collab}}\begin{sloppypar}\hyphenpenalty=5000\widowpenalty=500\clubpenalty=5000\textbf{Yerevan Physics Institute,  Yerevan,  Armenia}\\*[0pt]
V.~Khachatryan, A.M.~Sirunyan, A.~Tumasyan
\vskip\cmsinstskip
\textbf{Institut f\"{u}r Hochenergiephysik der OeAW,  Wien,  Austria}\\*[0pt]
W.~Adam, T.~Bergauer, M.~Dragicevic, J.~Er\"{o}, C.~Fabjan, M.~Friedl, R.~Fr\"{u}hwirth, V.M.~Ghete, J.~Hammer\cmsAuthorMark{1}, S.~H\"{a}nsel, C.~Hartl, M.~Hoch, N.~H\"{o}rmann, J.~Hrubec, M.~Jeitler, G.~Kasieczka, W.~Kiesenhofer, M.~Krammer, D.~Liko, I.~Mikulec, M.~Pernicka, H.~Rohringer, R.~Sch\"{o}fbeck, J.~Strauss, A.~Taurok, F.~Teischinger, P.~Wagner, W.~Waltenberger, G.~Walzel, E.~Widl, C.-E.~Wulz
% ...
\vskip\cmsinstskip
% ...
1:~~Also at CERN, European Organization for Nuclear Research, Geneva, Switzerland\\
% ...

\end{sloppypar}
\end{document}